\documentclass[aps,prc,twocolumn,longbibliography,reprint,superscriptaddress,amsmath,nofootinbib]{revtex4-1}
\usepackage[colorlinks=true, urlcolor=blue,linkcolor=blue, anchorcolor=blue, citecolor=red]{hyperref}
\usepackage[utf8]{inputenc}
\usepackage{natbib}
\usepackage{amsmath,amsfonts,amsthm,amssymb}
\usepackage{afterpage}
\usepackage{mathrsfs}
\usepackage{graphicx}
\usepackage[dvipsnames]{xcolor}
\usepackage{bm}
\usepackage{subfigure}

\graphicspath{{Fig/}}

\newcommand{\nn}{\nonumber\\ }

\newcommand{\pder}[2]{\frac{\partial#1}{\partial#2}}
\def\tR{\tau_R}
\def\p{{\boldsymbol p}}
\def\L{{\cal L}}
\def\M{{\cal M}}
\newcommand{\rmd}{{\mathrm d}}
\newcommand{\del}{\partial}
\def\fin{{f_{\mathrm{in}}}} 
\def\feq{{f_{\mathrm{eq}}}}
\newcommand{\fs}{{\rm fs}}

\begin{document}

\preprint{}

\title{From moments of the distribution function to hydrodynamics: The non-conformal case}


\author{Sunil Jaiswal}
\email{sunil.jaiswal@tifr.res.in}
\affiliation{Department of Nuclear and Atomic Physics, Tata Institute of Fundamental Research, Mumbai 400005, India}

\author{Jean-Paul Blaizot}
\email{jean-paul.blaizot@ipht.fr}
\affiliation{Institut de Physique Th{\'e}orique, Universit\'e Paris Saclay, CEA, CNRS, F-91191 Gif-sur-Yvette, France} 

\author{Rajeev S. Bhalerao}
\email{rajeev.bhalerao@iiserpune.ac.in}
\affiliation{Department of Physics, Indian Institute of Science Education and Research (IISER), Homi Bhabha Road, Pashan, Pune 411008, India}

\author{Zenan Chen} 
\email{19110200014@fudan.edu.cn}
\affiliation{Institute of Modern Physics Fudan University, 220 Handan Road, 200433, Yangpu District, Shanghai, China}

\author{Amaresh Jaiswal}
\email{a.jaiswal@niser.ac.in}
\affiliation{School of Physical Sciences, National Institute of Science Education and Research, An OCC of Homi Bhabha National Institute, Jatni-752050, India}

\author{Li Yan} 
\email{cliyan@fudan.edu.cn}
\affiliation{Institute of Modern Physics Fudan University, 220 Handan Road, 200433, Yangpu District, Shanghai, China}

\begin{abstract}

We study the one-dimensional boost-invariant Boltzmann equation in the relaxation-time approximation using special moments of the distribution function for a system with a finite particle mass. The infinite hierarchy of moments can be truncated by keeping only the three lowest moments that correspond to the three independent components of the energy-momentum tensor. We show that such a three-moment truncation reproduces accurately the exact solution of the kinetic equation after a simple renormalization that takes into account the effects of the neglected higher moments. We derive second-order Israel-Stewart hydrodynamic equations from the three-moment equations, and show that, for most physically relevant initial conditions, these equations yield results comparable to those of the three-moment truncation, albeit less accurate. We attribute this feature to the fact that the structure of Israel-Stewart equations is similar to that of the three-moment truncation. In particular, the presence of the relaxation term in the Israel-Stewart equations, yields an early-time regime that mimics approximately the collisionless regime. A detailed comparison of the three-moment truncation with second-order non-conformal hydrodynamics reveals ambiguities in the definition of second-order transport coefficients. These ambiguities affect the ability of Israel-Stewart hydrodynamics to reproduce results of kinetic theory.

\end{abstract}

\maketitle


\section{Introduction} 
\label{sec:introduction}

In this paper, we consider a fluid of massive particles that mimics the matter created in the collision of heavy nuclei at high energy. The matter is rapidly expanding along the collision axis (the $z$-axis), and we assume that this longitudinal expansion is invariant under Lorentz boosts in the $z$-direction (so-called Bjorken symmetry \cite{Bjorken:1982qr}). We also assume that the matter distribution is uniform in the transverse directions, i.e., in the plane transverse to the collision axis. The problem is thus reduced to a 1+1 dimensional problem amenable to a very detailed analysis. Our main interest is to understand, within the framework of kinetic theory, the transition between the early-time behavior of the system, which is essentially collisionless, and the late time dominated by collisions that lead eventually to the hydrodynamic behavior. There have been recently many studies of similar systems (for recent reviews see \cite{Florkowski:2017olj, Romatschke:2017ejr, Soloviev:2021lhs} and references therein, and for works more directly related to the present study see \cite{Heller:2015dha, Blaizot:2017lht, Blaizot:2017ucy, Strickland:2018ayk, Behtash:2019txb, Jaiswal:2019cju, Kurkela:2019set, Blaizot:2019scw, Blaizot:2020gql, Blaizot:2021cdv, Ambrus:2021sjg, Chattopadhyay:2021ive, Jaiswal:2021uvv}). This paper may be seen as an extension, to the case of massive particles, of a series of works on systems of massless particles by two of the authors of the present paper \cite{Blaizot:2017lht, Blaizot:2017ucy, Blaizot:2019scw, Blaizot:2021cdv}. It was shown in particular that, in the massless case, the dynamics is controlled by two fixed points of a nonlinear differential equation, each fixed point being associated to a well identified dynamics: a fixed point characterizing the collisionless, free-streaming motion of the particles, and the hydrodynamic fixed point. In fact, as we shall recall in this paper, the collisionless regime is itself governed by two fixed points, referred to as stable and unstable. The relevant one is the stable fixed point, which under the collisions, evolves slowly towards the hydrodynamic fixed point at late time. As in \cite{Blaizot:2021cdv}, we define the ``attractor solution'' as the particular solution that joins these two fixed points. This solution represents approximately the evolution of the stable free-streaming fixed point under the effect of the collisions. All solutions eventually converge to this attractor, sooner or later, depending on the initial conditions. The vicinity of each of the two fixed points is easily described either by a perturbation theory near the collisionless fixed point, or the Navier-Stokes hydrodynamics near the hydrodynamic fixed point. It is important to realise that, as emphasized in \cite{Blaizot:2021cdv}, Israel-Stewart-like conformal hydrodynamic theories capture the influence of the collisionless fixed point, albeit only approximately. This is the reason why Israel-Stewart like theories, when applied to Bjorken flow, seemingly work well even when extended into the collisionless regime, that is beyond the range of validity of hydrodynamics. In fact, it was found that a simple renormalization of a second-order transport coefficient in such theories is sufficient to make them capable of reproducing accurately the results of kinetic theory \cite{Blaizot:2021cdv}. 

The analysis of \cite{Blaizot:2017lht, Blaizot:2017ucy, Blaizot:2019scw, Blaizot:2021cdv} is based on special moments that take into account the angular distortions of the momentum distribution. These moments do not allow to reproduce the details of the momentum distribution, but are sufficient to account for the independent components of the energy-momentum tensor. In the case of massless particles, the energy-momentum tensor has only two components, the energy density and the difference between the longitudinal and the transverse pressures. When particles are massive, an additional degree of freedom appears, which is the trace of the energy-momentum tensor. We need therefore three moments to describe the system, three moments that turn out to be simple linear combinations of hydrodynamic fields, namely the energy density, and the shear and bulk pressures. Although these moments obey equations that belong to infinite hierarchies of coupled equations, as was shown in \cite{Blaizot:2019scw}, the truncation that consists in keeping only those equations that involve the hydrodynamic fields are sufficient to obtain a quantitative description of the evolution of the system. This is due to the specific fixed-point structure of the collisionless regime that we have recalled earlier, which is to some extent reproduced in Israel-Stewart hydrodynamics. As we shall see in this paper, the fixed-point structure identified in the collisionless regime for the moments is unaffected by the mass%
    \footnote{The fixed-point structure that emerges in hydrodynamic equations \cite{Jaiswal:2021uvv} does not agree with the fixed-point structure in kinetic theory for the choice of the second-order transport coefficients obtained in \cite{Denicol:2014vaa, Jaiswal:2014isa}. See Sections~\ref{sec_trunc_FP} and \ref{sec:newcoeff}, and Ref.~\cite{SunilJPB} for discussion on this issue.}.

This paper bears some similarity with Refs.~\cite{Romatschke:2017acs, Chattopadhyay:2021ive, Jaiswal:2021uvv, Chen:2021wwh, Kamata:2022jrc}. Our analysis confirms some of the conclusions of \cite{Chattopadhyay:2021ive, Jaiswal:2021uvv} where an `early-time attractor' was shown to emerge in the longitudinal pressure in kinetic theory, driven by the fast longitudinal expansion. This feature emerges naturally from our moment equations, as a consequence of a specific  decoupling of one set of moments from the other set. We discuss this in Section~\ref{sec_FP_ln} of the present paper, where the fixed-point structure of the moment set associated with the longitudinal pressure is analyzed. However, we differ in our analysis from Refs.~\cite{Chen:2021wwh, Kamata:2022jrc}, where fixed points are looked for within the second-order hydrodynamic flows. In our analysis, the essential fixed points are associated with distinct physical regimes, the collisionless and the hydrodynamic regimes, and we are mainly concerned with the transition from one regime to the other. 

This paper is organised as follows. The next Section~\ref{sec:kinetic} reviews the basic formalism. We extend the definition of the moments of the distribution function that were introduced in the massless case, and introduce a new set of moments needed to account for the non-vanishing trace of the energy-momentum tensor in the massive case. Equations of motion for these moments are obtained in the relaxation-time approximation for the collision integral. In the following short Section~\ref{sec:FS_regime} we show how generic properties of the collisionless regime are recovered in terms of moments, and provide a first indication on the free-streaming fixed points that play an important role in our analysis. The Section~\ref{sec_trunc_FP} is the main section of the paper. There we show how the main results of the massless case can be extended to the massive case. In particular, we consider a simple three-moment truncation where one keeps only those moments that identify with hydrodynamic fields. We show how  a simple renormalization of these three-moment equations allows us to reproduce accurately the attractor of the exact kinetic equation. In Section~\ref{sec:mom_hydro}, we show that the three-moment equations contain viscous hydrodynamics as a limiting case and we match the results to those of second-order hydrodynamics via a simple implementation of a Chapman-Enskog expansion. We underline ambiguities that arise in defining transport coefficients of second-order hydrodynamics. The final Section~\ref{sec_conclusions} summarises our conclusions. Appendix~\ref{app:exactKT_sol} contains details on the numerical implementation. The roles of the mass and the anisotropy of the initial distribution are discussed in Appendix~\ref{app:B}. We provide the details on the Chapman-Enskog expansion in Appendix~\ref{app:CE_exp}. The coefficients of gradient series generated from moment equations and from second-order hydrodynamics are compared in Appendix~\ref{app:gr}.

\vspace{-.2cm}
\section{Kinetic description of gas of massive particles} 
\label{sec:kinetic}
\vspace{-.2cm}

We consider a fluid of particles of mass $m$ undergoing Bjorken flow  along the $z$ direction \cite{Bjorken:1982qr}. The symmetry of the system implies that the single-particle distribution function $f$ depends on space-time solely through the proper time $\tau=\sqrt{t^2-z^2}$, and obeys the kinetic equation \cite{Baym:1984np}
\begin{equation}\label{keq_1}
\left(\frac{\partial}{\partial \tau} - \frac{p_z}{\tau} \frac{\partial}{\partial p_z}\right) f(\tau, p)= \mathcal{C}[f(\tau, p)],
\end{equation}
where $\mathcal{C}[f]$ denotes the collision integral. In the relaxation-time approximation (RTA) Eq.~(\ref{keq_1}) becomes
\begin{equation}\label{keq_2}
\left(\frac{\partial}{\partial\tau} - \frac{p_z}{\tau} \frac{\partial}{\partial p_z}\right)f(\tau, p) = -\frac{f(\tau, p)- f_{\rm eq}(p_0/T)}{\tR},
\end{equation}
where $\tau_R$ denotes the relaxation time. In the present work, we assume that $\tau_R$ is constant, independent of time.

The equilibrium distribution in the r.h.s. of Eq.~\eqref{keq_2} is a function of $p_0/T$, where  $p_0=\sqrt{m^2+p^2}$ is the energy of a particle with momentum $p$, and the temperature $T$ is determined from the requirement that the energy density be the same whether calculated with the true distribution function $f(\tau,p)$ or with the equilibrium distribution $f_{\rm eq}(p_0/T)$ (Landau matching condition). In this work, we do not introduce any chemical potential that could be associated to a conserved current. 

The solution of the kinetic Eq.~\eqref{keq_2} exhibits generically two regimes: a collisionless regime when $\tau\ll \tau_R$ and a collision-dominated (or hydrodynamic) regime at late times, $\tau\gg\tau_R$. When the mass is chosen to be constant, as it is in the present paper, an unphysical regime may be reached at late time. Indeed the equilibrium distribution function $f_{\rm eq}$, as well as all thermodynamic functions, are functions of the ratio $m/T$. Since the temperature decreases as the system expands, this ratio becomes eventually very large, and the particles become effectively non-relativistic. One could correct for this unsatisfactory behavior by introducing a thermal mass  proportional to the temperature \cite{Gorenstein:1995vm, Sasaki:2008fg, Bluhm:2010qf, Romatschke:2011qp, Czajka:2017wdo, Tinti:2016bav}. This is however beyond the scope of the present paper. We shall explore here only situations where this peculiar regime is pushed to very late times, of no relevance to the physics that we want to discuss%
    \footnote{A thorough mathematical discussion of this regime has been reported recently in \cite{Kamata:2022jrc}.}.

\vspace{-.2cm}
\subsection{General relations} 
\label{sec:general}
\vspace{-.2cm}

The energy-momentum tensor can be  obtained from the single-particle distribution function as%
	\footnote{Notation: $$ \int_\p \equiv \int \frac{d^3 p}{(2\pi)^3}.$$}
\begin{equation}\label{t_mu_nu}
T^{\mu\nu} = \int_\p \frac{p^\mu p^\nu}{p_0} f(\tau,p)\,.
\end{equation}
Because of the Bjorken symmetry, this tensor has only three independent components, the energy density $\varepsilon(\tau)$, the longitudinal pressure $P_L$, and the transverse pressure $P_T$, where 
\begin{align}\label{pl_pt}
\varepsilon &= \int_\p p_0 \, f(\tau, p),
\nn
P_L(\tau) &= \int_\p \frac{p_z^2}{p_0} \, f(\tau, p), 
\nn
P_T(\tau) &= \frac{1}{2} \int_\p \frac{p_\perp^2}{p_0} \, f(\tau, p).
\end{align}
The Landau matching condition mentioned above translates into the equation
\begin{equation}\label{keq_eng}
\varepsilon = \int_\p p_0 \, f(\tau, p) = \int_\p  p_0 \, \feq(p_0/T(\tau)),
\end{equation}
which determines the effective temperature $T(\tau)$ as a function of time. The trace of the energy-momentum tensor can be obtained by using the above equations:
\begin{align}\label{trace}
T^\mu_\mu =\varepsilon - P_L- 2P_T = m^2 \int_\p \frac{1}{p_0}f(\tau, p).
\end{align}
As the last equality shows, $T^\mu_\mu$ is non vanishing if $m\ne 0$. At equilibrium, $P_L=P_T=P$, where $P$ is the equilibrium pressure. The trace of the energy-momentum tensor is then $(T^\mu_\mu)_{\rm eq} = \varepsilon-3 P $.

Instead of using the variables $\varepsilon$, $P_L$, and $P_T$ to represent the energy-momentum tensor, it is customary, for systems close to local equilibrium, i.e., in the hydrodynamic regime, to use instead the bulk ($\Pi$) and shear ($\phi$) pressures. These are defined as follows
\begin{align}
P_T &= P+\Pi+\phi/2, \qquad\, P_L =P+\Pi-\phi ,
\nn
\phi &= \frac{2}{3}(P_T-P_L), \quad P+\Pi= \frac{1}{3} (P_L+2P_T),
\end{align}
where $P$, the equilibrium pressure, is related to the energy density by an equation of state, $P=P(\varepsilon)$. In this work we use the equation of state of an ideal classical gas (with Boltzmann statistics). In terms of the distribution function, these quantities are given by 
\begin{align}
P+\Pi &= \frac{1}{3}\int_\p \frac{p^2}{p_0} f(\tau,p),
\nn
\phi &=-\frac{2}{3} \int_\p \frac{1}{p_0} \left(p_z^2-\frac{p_\perp^2}{2}  \right)f(\tau,p),
\end{align}
with 
\begin{align}
P &= \frac{1}{3}\int_\p \frac{p^2}{p_0} f_{\rm eq}(p_0/T),
\nn
\Pi &= \frac{1}{3}\int_\p \frac{p^2}{p_0} \left(f(\tau,p)-f_{\rm eq}(p_0/T)\right).
\end{align}

\vspace{-.2cm}
\subsection{Sets of special moments} 
\label{sec:kinetic_moments}
\vspace{-.2cm}

One can also express $\varepsilon, P_L, P_T$ in terms of $\L_n$-moments, defined generally by \cite{Blaizot:2017lht,Blaizot:2017ucy}
\begin{equation}\label{L_mom}
\L_n \equiv  \int_\p p_0 \ P_{2n}(\cos \psi) \ f(\tau, p), 
\qquad \forall n \geq 0
\end{equation}
where $\cos\psi \equiv p_z/p_0=v_z$ is the velocity of a constituent projected along the $z$-axis, and $P_{2n}$ is the Legendre polynomial of order $2n$. Note that the integrand is even in $p_z \rightarrow -p_z.$ The first two $\L_n$-moments are
\begin{subequations}\label{eq:L0L1}
\begin{align}
\label{L_0}
\L_0 &= \int_\p p_0  \ f(\tau, p) = \varepsilon \,,
\\ \label{L_1}
\L_1 &=  \int_\p p_0 \ P_{2}(\cos \psi) \ f(\tau, p) \nn
	 &= \frac{1}{2} \int_\p  \frac{1}{p_0}(3p_z^2 -p_0^2) \ f(\tau, p)
=\frac{3}{2}P_L -\frac{\varepsilon}{2} \,.
\end{align}
\end{subequations}
As is clear from these equations, the energy density $\varepsilon$ and the longitudinal pressure $P_L$ can be expressed in terms of these two moments
\begin{equation}\label{hydro_mom}
\varepsilon = \L_0 , \qquad P_L = \frac{1}{3} \left(\L_0 + 2 \L_1 \right). 
\end{equation}
The transverse pressure involves in addition the trace of the energy-momentum tensor
\begin{equation}\label{transversepressure}
P_T = \frac{1}{3} \left( \L_0 -\L_1 -\frac{3}{2} T^\mu_\mu \right),
\end{equation}
which cannot be expressed solely in terms of the $\L_n$-moments. This  involves another type of moments, which we define as
\begin{equation}\label{M_mom}
\M_n \equiv m^2 \int_\p \frac{1}{p_0} P_{2n}(\cos \psi) \ f(\tau, p), 
	 \qquad \forall n \geq 0.
\end{equation}
These moments $\M_n$ vanish when $m=0$. The moment $\M_0$ is equal to the trace of the energy-momentum tensor:
\begin{equation}\label{M_0}
\M_0 = m^2\int_\p \frac{1}{p_0} f(\tau, p) = T^\mu_\mu = \varepsilon-P_L-2 P_T.
\end{equation}
It follows that the fluid energy-momentum tensor of systems undergoing Bjorken flow can be written in terms of the three moments $\L_0, \L_1$, and $\M_0$. Note that the equilibrium value of the $\M_0$-moment is
\begin{equation}
\label{M0_eq}
\M_0^{\rm eq} = (T^\mu_\mu)_{\rm eq} = -2 \L_1^{\rm eq} = \varepsilon-3 P \neq 0.
\end{equation}
Note also the relations  
\begin{equation}
\label{eq_isoP_phi}
P+\Pi = \frac{1}{3} \left( \L_0 - \M_0 \right), \quad 
\phi = -\frac{2}{3} \left( \L_1 + \frac{\M_0}{2} \right) \,,
\end{equation}
and more generally 
\begin{equation}
\L_n-{\M}_n = \int_\p \frac{p^2}{p_0} P_{2n}(\cos\psi)\, f(\tau,p).
\end{equation}

In the massless case, the moments $\L_n$ with $n\ge 1$ vanish for a spherically symmetric $f(\tau,p)$. They thus have a simple geometrical interpretation, as their values reflect the deviation of the momentum distribution from a spherical shape. This is no longer the case in the massive case, as the moments $\L_n$ acquire a nonzero value even for a spherical distribution, such as the equilibrium distribution function. This may be traced back to the fact that, in the definition \eqref{L_mom}, $\cos\psi$ denotes the projection of the velocity on the $z$ axis, and this takes values smaller than one when $m\ne 0$. Thus the angular integration is not ``complete'' (the complete momentum integration would involve $\cos\theta=p_z/|\p|$ instead of $\cos\psi=p_z/p_0=v_z$). To illustrate this point, consider the moment $\L_1$. We have, starting from the definition \eqref{L_mom}, 
\begin{align}
\L_1 &= \int_\p p_0 \ P_{2}(\cos \psi) \ f(\tau, p)
\nn
	 &= \frac{1}{2} \int_\p  p_0 \left(3\frac{p_z^2}{p_0^2} -1\right) \ f(\tau, p)
\nn
	 &= \frac{1}{2} \int_\p  p_0 \left(3\frac{p_z^2}{p^2}\left(1-\frac{m^2}{p_0^2}\right) -1\right) \ f(\tau, p)
\nn
	 &= \int_\p  p_0 P_2(\cos\theta) \ f(\tau, p)-\frac{3m^2}{2} \int_\p  \frac{1}{p_0}\cos^2\theta \ f(\tau, p)
\nn
	 &= \hat \L_1-\frac{3m^2}{2} \int_\p  \frac{1}{p_0}\cos^2\theta \ f(\tau, p).
\end{align}
In going from the first line to the last, we have moved from an integration over $\psi$ to an integration over $\theta$. Clearly the moment $\hat \L_1$ in the last line vanishes for a spherical distribution: it plays the same role as the moment $\L_1$ in the massless case, as measuring the angular distortion of the momentum (velocity) distribution. In equilibrium, this contribution vanishes. The entire contribution to $\L_1$ comes then from the second term and we recover the result $\L_1^{\rm eq}= -(1/2) \M_0^{\rm eq}$ already mentioned in Eq.~(\ref{M0_eq}). The choice of the moments $\L_n$'s, as defined in Eq.~(\ref{L_mom}), is motivated by the fact that these moments obey nearly the same equations in the massless and the massive cases, as we shall see shortly. 

In kinetic theory, the positivity of the pressures and of the trace of the energy-momentum tensor entails several bounds on the various moments \cite{Chattopadhyay:2021ive, Jaiswal:2021uvv}:
\begin{align}\label{bounds}
\frac{\L_1}{\L_0} +\frac{3}{2} \frac{\M_0}{\L_0} \leq 1 \,, \quad
\frac{\L_1}{\L_0} \geq -\frac{1}{2} \,, \quad
0\le \frac{\M_0}{\L_0} \leq 1 \,.
\end{align}
These bounds should be satisfied, in particular, by the initial conditions. They should also be preserved during the evolution. However, we shall see that some of them may be violated when approximations are made, in particular in some truncation of the moment equations, or in the context of Israel-Stewart hydrodynamics.

\vspace{-.2cm}
\subsection{Equations for the \texorpdfstring{$\L$}{}-moments and the \texorpdfstring{$\M$}{}-moments} 
\label{sec:kinetic_eq}
\vspace{-.2cm}

By using well-known relations among the Legendre polynomials and the definition~\eqref{L_mom} of $\L$-moments, one can recast Eq.~\eqref{keq_2} into the following hierarchy of coupled equations for the $\L$-moments \cite{Blaizot:2017ucy}:
\begin{subequations}\label{L_eqn}
\begin{align}
\label{L_eqn_a}
\pder{\L_0}{\tau} =& -\frac{1}{\tau} \left( a_0 \L_0 + c_0 \L_1 \right) \,,\\
\label{L_eqn_b}
\pder{\L_n}{\tau} =& -\frac{1}{\tau} \left( a_n \L_n + b_n \L_{n-1} + c_n \L_{n+1} \right) 
	- \frac{\left( \L_n - \L_n^{\rm eq} \right)}{\tR} \,,
\end{align}
\end{subequations}
where the coefficients $a_n,b_n,c_n$ are pure numbers given by 
\begin{align}\label{L_coeff}
a_n &= \frac{2(14 n^2+7n-2)}{(4n-1)(4n+3)} \,, 
\nn
b_n &= \frac{(2n-1)2n(2n+2)}{(4n-1)(4n+1)} \,, 
\nn
c_n &= \frac{(1-2n)(2n+1)(2n+2)}{(4n+1)(4n+3)} \,.
\end{align}
These coefficients satisfy two important relations, valid for any $n$
\begin{subequations}\label{L_n:coeff_rel}
\begin{align}
\label{L_n:coeff_rel:1}
& a_n+b_n+c_n = 2 \,,
\\ \label{L_n:coeff_rel:2}
& a_n A_n +b_n A_{n-1} +c_n A_{n+1} = A_n \,,
\end{align}
\end{subequations}
where
\begin{equation}\label{An_def}
A_n = P_{2n}(0) =(-1)^n \frac{(2n-1)!!}{(2n)!!} \,.
\end{equation}
Note that $a_0=4/3,\, b_0=0,\, c_0=2/3$, and $A_0=1$. As we shall see in the next section, these relations determine the behaviors of the moments at late and early times in the collisionless regime. 

In deriving the equation for $\L_0$ in Eq.~\eqref{L_eqn_a}, we used the Landau matching condition, Eq.~\eqref{keq_eng}, to eliminate the collision term $\propto \L_0-\L_0^{\rm eq}$. The equation for ${\cal L}_n \, (n \ge 1)$ differs from the corresponding equation in the massless case, by the presence of the term $\L_n^{\rm eq}\ne 0$ in its right-hand side, which forces the moment $\L_n$ to relax to its equilibrium value $\L_n^{\rm eq}$. This equilibrium value can be computed using the equilibrium distribution, which is known once the temperature, or equivalently the energy density, is known. This introduces an explicit nonlinearity in the equations, since all the equilibrium moments depend implicitly on $\L_0$. However, the equations for the $\L_n$-moments constitute a closed system of equations that can be solved, independently of the $\M$ moments, in order to obtain in particular the evolution of the energy density ($\varepsilon$) and of the longitudinal pressure ($P_L$) using Eq.~\eqref{hydro_mom}. 

In order to obtain the evolution for the transverse pressure $P_T$, we need the evolution of $\M_0$, Eq.~\eqref{M_0}. This equation is part of an infinite set of  coupled equations for the $\M$-moments which can be derived in the same way as those for the $\L$-moments, starting from Eqs.~\eqref{keq_2} and using the definition \eqref{M_mom}. One gets
\begin{equation}
\label{M_eqn}
\pder{\M_n}{\tau} \!=\! -\frac{1}{\tau} \left(a'_n \M_n + b'_n \M_{n-1} + c'_n \M_{n+1}\right) 
	- \frac{( \M_n \!-\! \M_n^{\rm eq} )}{\tR}  \,,
\end{equation}
Here the coefficients $a'_n,b'_n,c'_n$ are real constants given by
\begin{align}
a'_n &= \frac{2(6n^2+3n-1)}{(4n-1)(4n+3)} \,,
\nn
b'_n &= \frac{4n^2(2n-1)}{(4n-1)(4n+1)} \,,
\nn
c'_n &= -\frac{(2n+1)^2(2n+2)}{(4n+1)(4n+3)} \,.
\end{align}
They satisfy relations analogous to the relations (\ref{L_coeff}) and valid for any $n$,
\begin{subequations}\label{M_n:coeff_rel}
\begin{align}
\label{M_n:coeff_rel:1}
& a'_n+b'_n+c'_n = 0 \,,
\\ \label{M_n:coeff_rel:2}
& a'_n A_n +b'_n A_{n-1} +c'_n A_{n+1} = A_n \,.  
\end{align}
\end{subequations}
Note that in Eq.~(\ref{M_n:coeff_rel:2}), the coefficients $A_n$ are the same as in Eq.~(\ref{An_def}), in spite of the fact that $a_n', b_n', c_n'$ differ from $a_n, b_n, c_n$. 

In the collisionless regime ($\tau_R\to\infty$) the equations for the $\M$-moments decouple from those of the $\L$-moments. The collisions introduce a coupling between the two sets of moments, via the presence of the equilibrium moments $\M_n^{\rm eq}$ in the right-hand side of Eq.~(\ref{M_eqn}), which depend on the energy density. Thus, in the presence of collisions,  the equations for the $\M$-moments can only be solved after one has solved the equation for the $\L$-moments, in order to obtain  in particular the energy density. 

Note that the mass does not appear explicitly in any of these moment equations. The mass enters only through the equilibrium distribution, i.e., the equilibrium values of the moments, e.g., $\L_1^{\rm eq}=-(1/2)(\varepsilon-3P)$,  and through the initial conditions (see the discussion in the next section and in  Appendix~\ref{app:exactKT_sol}).

\vspace{-.2cm}
\section{The collisionless regime}
\label{sec:FS_regime}
\vspace{-.2cm}

At this point, we examine how the equations for the moments describe the collisionless regime. In particular, we recall the observation, thoroughly discussed in \cite{Blaizot:2019scw} in the case of massless particles, that   simple truncations of the $\L$-moment equations capture the most important aspects of the dynamics of the lowest moments. Since, as we have already emphasized, the equations for the moments are independent of the mass of the particles, this property remains true in the present case of massive particles. It is connected to a specific fixed-point structure of the operators in the left-hand side of Eq.~(\ref{keq_1}) that describes the collisionless regime. 

Observe first that the behavior of the moments in the free-streaming regime can be determined directly from  the solution of Eq.~\eqref{keq_1} in which one ignores the collision term in the right-hand side (or equivalently let $\tau_R\to\infty$). Let $\fin(p_\perp,p_z)$ denote the distribution function at some initial proper time $\tau_{\rm in}$. Without any loss in generality%
    \footnote{In the calculation of attractor solutions in the next section, we shall be led to consider anisotropic initial distributions. These may be seen as  resulting from free-streaming evolution from an earlier time of isotropic distributions, or equivalently from squeezing the longitudinal momentum by a scale factor $\xi$ playing the same role as $\tau/\tau_{\rm in}$, namely choosing $f_{\rm in}(p_0)=f_{\rm in}(p_\perp,p_z \xi)$.} 
we may assume that this initial distribution is isotropic, i.e., $\fin(p_\perp,p_z)$ depends only on $p_0=\sqrt{p_\perp^2+p_z^2+m^2}$. The solution of the collisionless equations at any time $\tau$ takes then the form  $f_\fs(\tau; p_\perp, p_z) = \fin(p_\perp, p_z (\tau/\tau_{\rm in}))$ \cite{Baym:1984np}. Thus, at late time, $\tau \gg \tau_{\rm in}$, the distribution becomes peaked around $p_z=0$. On the other hand, at early time, $\tau \ll \tau_{\rm in}$, the distribution becomes elongated along the $z$ axis. These two solutions will be associated later with two fixed points of the free-streaming motion. 

A simple calculation allows us to determine the leading-order behaviors of the moments at early and late times, starting from an isotropic initial distribution, as just mentioned. To be specific, let us consider the energy density. This is given by
\begin{equation}
\varepsilon(\tau)=\int_\p \left( m^2+p_\perp^2+p_z^2 \right)^{1/2} \fin(p_\perp, p_z (\tau/\tau_{\rm in})), 
\end{equation}
or, after a simple change of variable, by
\begin{equation}\label{epsilontau}
\varepsilon(\tau)=\frac{\tau_{\rm in}}{\tau}\int_\p \left( m^2+p_\perp^2+p_z^2 (\tau_{\rm in}^2/\tau^2) \right)^{1/2} \fin(p_\perp, p_z).
\end{equation}
It is then straightforward to show that the leading behaviors of $\varepsilon(\tau)$ at late and early times are, respectively, 
\begin{equation}
\varepsilon(\tau\gg \tau_{\rm in})\propto \frac{1}{\tau},\qquad 
\varepsilon(\tau\ll \tau_{\rm in})\propto \frac{1}{\tau^2}.
\end{equation}
In the late-time regime, the factor $1/\tau$ reflects the expansion, the excitations carrying typical energy $\approx \sqrt{p_\perp^2+m^2}$. The extra factor of $1/\tau$ at early time originates from the fact that the energy of the excitations is dominated by their longitudinal momentum that scales as $1/\tau$. 

It is not difficult to see that the same reasoning leads to the conclusion that all the $\L_n$ moments share the same behavior as that of the energy density at early and late times, namely
\begin{equation}\label{largetauL}
\L_n(\tau\gg \tau_{\rm in})\propto \frac{1}{\tau},\qquad \L_n(\tau\ll \tau_{\rm in})\propto \frac{1}{\tau^2}.
\end{equation}
When we apply the same analysis to the $\M$-moments, one finds instead the behaviors
\begin{equation}\label{largetauM}
\M_n(\tau\gg \tau_{\rm in})\propto \frac{1}{\tau},\qquad \M_n(\tau\ll \tau_{\rm in})\propto {\rm const.}
\end{equation}

As we show now, the same behaviors are recovered from a simple analysis of the coupled equations for the moments. To proceed, we rewrite the  equations for the moments as linear homogeneous matrix equations (with here $t\equiv\ln (\tau/\tau_{\rm in})$) 
\begin{equation}\label{matrix_eqn}
\partial_t \Vec{\L} = - G \Vec{\L}, \qquad \partial_t \Vec{\M} = - H \Vec{\M},
\end{equation}
where $\Vec{\L} \text{ and } \Vec{\M}$ are vectors in the infinite-dimensional spaces spanned by the $\L_n$ and $\M_n$ moments, respectively, and $G, H$ are tridiagonal matrices with constant elements. 

The equations for the $\L_n$ moments are identical to those of the massless case and have been studied extensively in \cite{Blaizot:2019scw}. Let us recall that there are two (and only two) real eigenvalues, already well approximated by a two-moment truncation (i.e., setting $\L_2=0$ in the equation (\ref{L_eqn_b}) for $n=1$), all other eigenvalues being complex. The eigenvalues of the two moment truncation are  $\lambda_0 \simeq 0.93$ and $\lambda_1 \simeq 2.21$.  These eigenvalues converge slowly to their exact values $(\lambda_0 \to 1, \ \lambda_1 \to 2)$ as one increases the level of truncation. The eigenvector $\Vec{\L}$ corresponding to the eigenvalue ($\lambda_0=1$) has all its components proportional, $\L_n=A_n \L_0$. That this corresponds to a solution can be seen immediately by substituting  $\L_n\mapsto A_n \L_0$ in Eq.~(\ref{L_eqn_b}) and using the  relation~\eqref{L_n:coeff_rel:2}. This solution, in which all moments decay as $1/\tau$, represents the generic behavior of the free-streaming moments at late time, as discussed above. This behavior will be referred to, in the next section, as the {\it stable free-streaming fixed point}. The components of the eigenvector $\Vec{\L}$ corresponding to the eigenvalue ($\lambda_1=2$) are $\L_n=\L_0$. This solution follows from the relation~\eqref{L_n:coeff_rel:1}, and corresponds to the situation where all the moments decay as $1/\tau^2$. This behavior will be referred to, in the next section, as the {\it unstable free-streaming fixed point}. 

The equations for the $\M_n$ moments have a simpler structure. They have only one real eigenvalue, which, in the lowest-order truncation, i.e., keeping only the moment $\M_0$, is $\mu_0 = 2/3$. Truncation at higher orders introduces new complex conjugate eigenvalues, with the real eigenvalue $\mu_0$ appearing at every even truncation (truncations at odd $n$ give only complex eigenvalues), eventually converging to $\mu_0 =1$. The convergence is very slow; e.g., when truncating the equations at  $n=10$,  $\mu_0\simeq 0.7$, still far from 1. The components of the eigenvector corresponding to the eigenvalue ($\mu_0=1$) satisfy $\M_n=A_n \M_0$, as can be verified from the relation~\eqref{M_n:coeff_rel:2}, and all moments decay as $1/\tau$ at late time. There exists another relation~\eqref{M_n:coeff_rel:1} between the coefficients of $\M_n$-moment equations. Using this relation in Eqs.~\eqref{M_eqn}, we obtain a special solution in which  all the $\M_n$-moments are equal and constant in time.

The particular structure of the moment equations for the free-streaming evolution that we have briefly recalled, with in particular the existence of simple fixed points whose physical interpretation is transparent, plays an essential role in the foregoing discussion. Note that the notion of fixed point used here refers to the asymptotic time dependence of the free-streaming moments at late time, given by a power law independent of the mass. The absolute values of the moments at late time depends on the initial condition, and therefore on the mass. This can be clearly seen for instance  from the expression (\ref{epsilontau}) of the energy density: at late time, $\varepsilon(\tau)\propto 1/\tau$ where the proportionality coefficient is given by an integral that clearly depends on the mass.

\vspace{-.2cm}
\section{Truncation and fixed-point analysis}
\label{sec_trunc_FP}
\vspace{-.2cm}

We shall now consider the equations for the three moments $\L_0$, $\L_1$, and $\M_0$ which fully represent the energy-momentum tensor. We rewrite these equations here for convenience
\begin{subequations}
\label{L_n:trunc}
\begin{align}
\label{eq_L_0}
\pder{\L_0}{\tau} &= -\frac{1}{\tau} \left( a_0 \L_0 + c_0 \L_1 \right) , 
\\ \label{eq_L_1}
\pder{\L_1}{\tau} &= -\frac{1}{\tau} \left( a_1 \L_1 + b_1 \L_0 +c_1 \L_2  \right) 
	- \frac{\left( \L_1 - \L_1^{\rm eq} \right)}{\tR}   ,
\\ \label{eq_M0}
\!\!\pder{\M_0}{\tau} &= - \frac{1}{\tau}  \left( a'_0 \M_0 + c_0' \M_1 \right) - \frac{\left( \M_0 - \M_0^{\rm eq} \right)}{\tR}  \,,
\end{align}
\end{subequations}
where $\ a_0=4/3,\ c_0=2/3,\ a_1= 38/21,\ b_1=8/15,\ c_1=-12/35,\ a'_0=2/3$, and $c_0'=-2/3$. Note that these equations for $\L_0,\L_1$, and $\M_0$ are exact equations. But in order to solve them we need to know the two moments $\L_2$ and $\M_1$. We shall discuss later convenient approximations for these two moments. For brevity, we shall often refer to the set of equations (\ref{L_n:trunc}) as the {\it three-moment truncation}. This includes the naive truncation where we set simply ${\cal L}_2=0$ and ${\cal M}_1=0$, or more sophisticated approximations where we use easily available information on $\L_2$ and $\M_1$ to construct more accurate approximations. 

As we have observed earlier, the equations for the $\L$-moments decouple from those of the $\M$-moments, and they can be solved independently. We shall then first analyze the solution of Eqs.~(\ref{eq_L_0}) and (\ref{eq_L_1}), and then proceed to the solution of Eq.~(\ref{eq_M0}) for the moment $\M_0$. 

\vspace{-.2cm}
\subsection{The coupled equations for the moments \texorpdfstring{$\L_0$}{} and \texorpdfstring{$\L_1$}{}}
\label{sec_FP_ln}
\vspace{-.2cm}

\subsubsection{The equivalent nonlinear equation and its fixed points}
\vspace{-.2cm}

We start by reformulating the eigenvalue problem for the moments $\L_0$ and $\L_1$ as an equivalent nonlinear equation, and then proceed to a fixed-point analysis \cite{Blaizot:2019scw}. Starting from Eqs.~(\ref{eq_L_0}) and (\ref{eq_L_1}), we eliminate the moment $\L_1$ and get the following equation for the energy density $\L_0$: 
\begin{align}\label{eqL0}
\tau\ddot\L_0 &+\left[1+a_0+a_1+\frac{\tau}{\tau_R} \right]\dot\L_0 + \left[a_0a_1-c_0b_1+a_0\frac{\tau}{\tau_R} \right]\frac{\L_0}{\tau}
\nn
&-c_0 c_1 \frac{\L_2}{\tau} =\frac{1}{2}\frac{c_0}{\tau_R}
(\varepsilon-3P),
\end{align}
where we have used $\L_1^{\rm eq}=-\frac{1}{2}(\varepsilon-3P)$ (see Eq.~\eqref{M0_eq}). We then transform this equation into an equation for the quantity
\begin{equation}\label{eq:g0def}
g_0=\frac{\tau}{\L_0} \frac{\del \L_0}{\del \tau},\qquad g_0=-a_0 -c_0 \frac{\L_1}{\L_0}.
\end{equation}
The left equation shows that, in the regimes where the moments behave as power laws,  $g_0$ is the exponent in that power law. The right equation, which follows immediately from Eq.~(\ref{eq_L_0}), shows that $g_0$ is directly related to the ratio $\L_1/\L_0$. In the massless case, this quantity measures the pressure asymmetry. As already mentioned, this is no longer the case in the massive case since the pressure asymmetry, which can be measured more generally by the shear pressure $\phi=(2/3) (P_T-P_L)$, differs from $\L_1$ (see Eq.~(\ref{eq_isoP_phi})). The determination of the pressure asymmetry requires, besides $\L_1$, also the estimate of the moment $\M_0$, which we shall come to later. 

When written in terms of $g_0$, Eq.~(\ref{eqL0}) becomes a nonlinear, first-order, differential equation that can be put in the following form
\begin{align}\label{eq:betafct}
w \frac{\rmd g_0}{\rmd w} =& \beta(g_0,w),\nn
-\beta(g_0,w) =& g_0^2+g_0\left [ a_0+a_1+w \right]+a_0a_1 -c_0b_1 +a_0 w
\nn
&-c_0 c_1 \frac{\L_2}{\L_0}-\frac{c_0}{2} w \left(1-3\frac{P}{\varepsilon}\right),
\end{align}
where  $w \equiv \tau/\tau_R$, and we recall that the relaxation time $\tau_R$ is taken to be constant in this paper. Since, in the massive case,  the pressure $P$ is a (known) function of $z\equiv m/T$, we need to follow the evolution of $z$ as a function of time. This is easily obtained from $g_0$ itself, since $\varepsilon=\L_0$ is a well defined function of $z$, i.e., $\L_0(\tau, z)=\L_0(z(\tau))$.  Using Eq.~\eqref{eq:g0def}, we get
\begin{equation}\label{eq:zevol_g0}
g_0= \frac{w}{\L_0} \frac{\partial \L_0}{\partial z} \frac{\partial z}{\partial w}
 \implies \frac{\partial z}{\partial w} = \frac{\L_0}{(\partial \L_0/\partial z)} \frac{g_0}{w} .
\end{equation}
The solution of these two coupled equations, Eqs.~(\ref{eq:betafct}) and (\ref{eq:zevol_g0}), is conveniently analyzed in terms of fixed points, that is, in terms of the zeroes of the function $\beta(g_0,w)$. 

At small time, $w \ll 1$, the function $\beta(g_0,w)$ is dominated by the terms that do not depend on $w$: the terms linear in $w$ can be ignored in leading order, so that in this regime the function  $\beta(g_0,w)$ is independent of $w$. The zeroes of this function correspond to the two fixed points  of the free-streaming evolution. The values of $g_0$ at these fixed points coincide (to within a sign) with the eigenvalues $\lambda_0$ and $\lambda_1$ mentioned in the previous section, and they are independent of the mass. However, these values depend on the moment $\L_2$. For instance, in the naive truncation where one simply ignores the term proportional to $\L_2$ in Eq.~(\ref{eq:betafct}), the fixed point values are respectively $-0.93$ and $-2.21$, as indicated in the previous section (note that these values are outside the bounds mentioned in Eq.~\eqref{bounds}). We shall return to the determination of $\L_2$ later in this section. 

\begin{figure*}[t!]
    \centering
    \includegraphics[width=\textwidth]{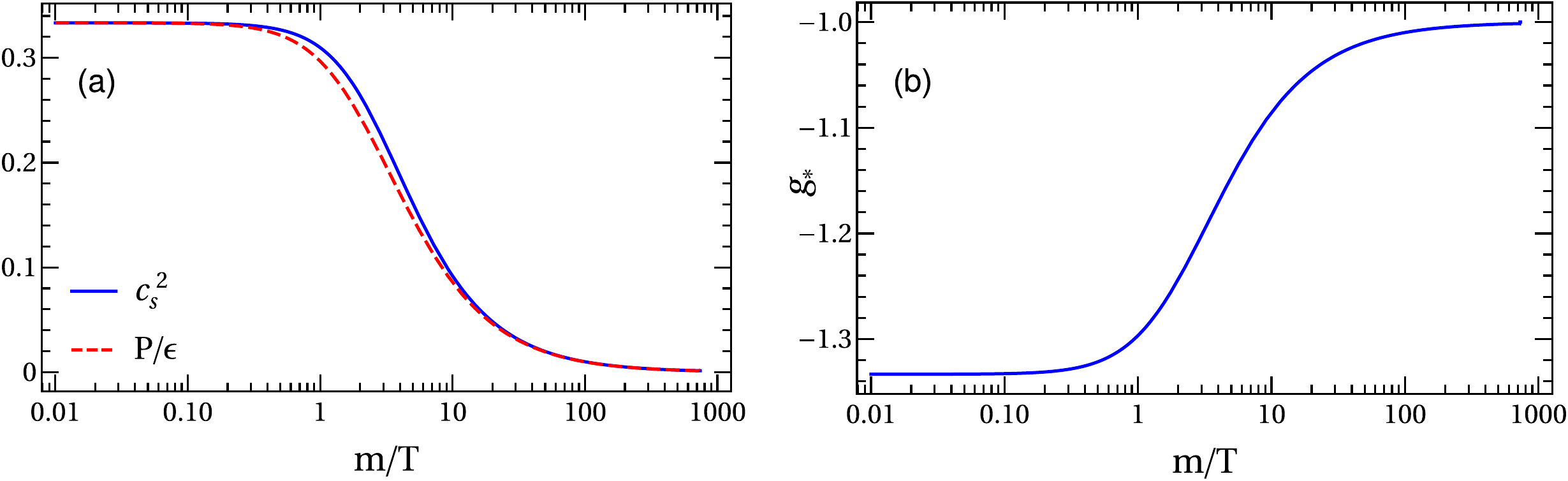}
    \vspace*{-6mm}
    \caption{The variation with $ m/T$ of $c_s^2$ and $P/\varepsilon$ (a) and of the hydrodynamic fixed point $g_*$ (b). In these calculations, we use the equation of state of an ideal gas of massive particles.}
    \vspace*{-2mm}
   \label{fig:cs2_gstar}
\end{figure*}

At late time, i.e., when $w\gg 1$, the dominant terms are the terms linear in $w$, whose cancellation determines the hydrodynamic fixed point $g_*$:
\begin{equation}\label{eq:hydrog*}
g_*+a_0 =\frac{c_0}{2}\left(1-3\frac{P}{\varepsilon}\right).
\end{equation}
In the massless case, the right-hand side vanishes, and the fixed point is given by $g_*=-a_0=-4/3$. In the general case, we have
\begin{equation}\label{eq:g*_fp}
g_*=-1-\frac{P}{\varepsilon}\approx -1-c_s^2,
\end{equation}
where we have used $c_0=2/3$,  $a_0-c_0/2=1$, and in the last step we have assumed that $P/\varepsilon\approx c_s^2$. That this is a good approximation can be inferred from Fig.~\ref{fig:cs2_gstar}(a) where the two quantities $P/\varepsilon$ and $c_s^2$ are compared. In the massless case, the hydrodynamic fixed point is a true fixed point, independent of $w$. In the massive case, this fixed point is a function of $m/T$, and therefore it evolves slowly with time, varying from from $g_* =-4/3$ as $m/T\to 0$ to $g_*=-1$ as $m/T\to \infty$, as shown in Fig.~\ref{fig:cs2_gstar}(b). Because of the expansion, the regime where  $m/T\to \infty$ is eventually reached at late time. As already discussed, this is not a physically interesting regime, and in the forthcoming numerical studies we shall restrict ourselves to cases where this regime is pushed to very late time, i.e., at $\tau\gg \tau_R$. 

\vspace{-.3cm}
\subsubsection{Approximate attractor solution}
\vspace{-.3cm}

\begin{figure*}[t!]
    \centering
    \includegraphics[width=\textwidth]{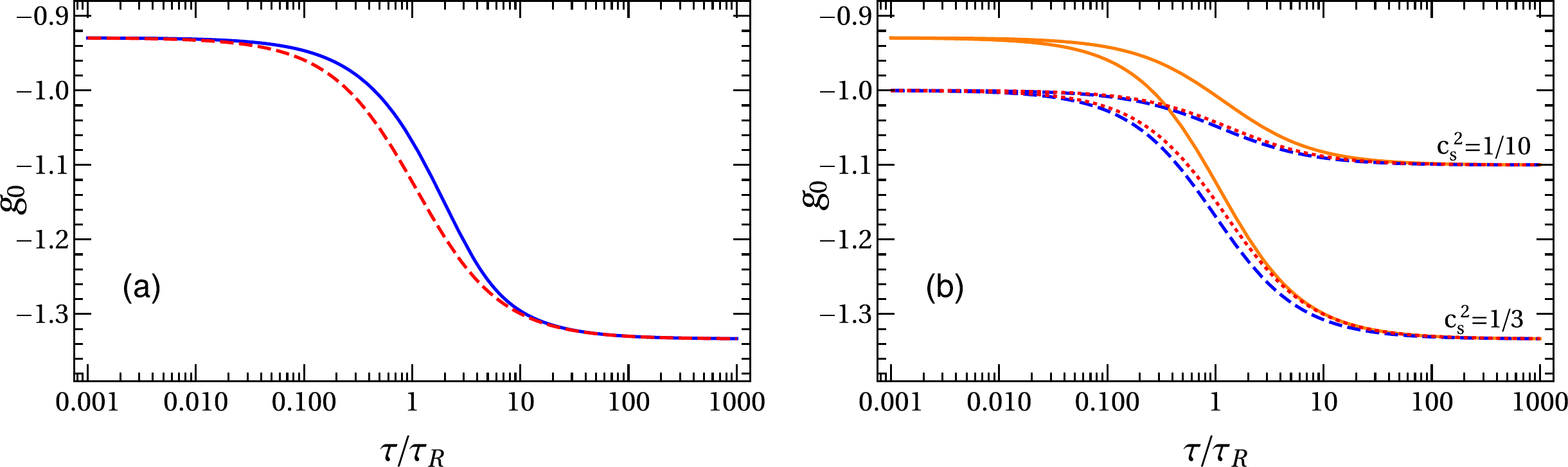}
    \vspace*{-6mm}
    \caption{{\bf (a)}: The attractor solution $g_0(w)$ as a function of $w\equiv\tau/\tau_R$, for massless particles and in the truncation $\L_2=0$. The (blue) solid curve represents the exact solution of Eqs.~(\ref{eq:betafct}) and (\ref{eq:zevol_g0}), while the (red) dashed curve is obtained by solving the implicit equation $\beta(g_0,w)=0$ rather than Eq.~(\ref{eq:betafct}). {\bf (b)}: Approximate attractor, obtained as the solution of $\beta(g_0,w)=0$ as a function of $w=\tau/\tau_R$, for two values of the speed of sound assumed to be constant, $c_s^2=1/3$ and $1/10$. The (orange) full lines are the solutions for $\L_2=0$, the (blue) dashed lines are obtained by setting $\L_2=A_2 \L_0$. The (red) dotted lines correspond to fixing $\L_2$ according to Eq.~(\ref{eq:a1prime}).}
    \vspace*{-2mm}
    \label{fig:g0_adiabatic}
\end{figure*}

The solution of Eq.~(\ref{eq:betafct}) that starts from the (stable) free-streaming fixed point at early times and reaches the hydrodynamic fixed point at late times is what we defined, in this context, as the attractor solution in \cite{Blaizot:2017ucy}. A good approximation of this solution can be obtained as the location of the zero of the function $\beta(g_0,w)$ as a function of $w$, that is, as the function $g_0(w)$ implicitly defined via the equation $\beta(g_0(w),w)=0$. This approximation, and its validity, are thoroughly discussed in \cite{Blaizot:2021cdv}, and its quality can be gauged from the plots in  Fig.~\ref{fig:g0_adiabatic}(a) corresponding to the massless case, and the naive truncation where $\L_2=0$. These attractor solutions are obtained  by solving  Eq.~(\ref{eq:betafct})  with initial value at $w_{\rm in} =10^{-3}$ equal to $g_0 \approx -0.92$ (which is the value of the stable free-streaming fixed point for $\L_2=0$). The small deviation between the exact and approximate solutions occurs when $\beta(g_0) = w \frac{\rmd g_0}{\rmd w}$ is significant, i.e., in the transition region that connects the free-streaming fixed point and the hydrodynamic fixed point.

The hydrodynamic fixed point (\ref{eq:hydrog*}) depends on the equation of state, but does not depend on the higher moments $\L_{n\ge 2}$. In contrast, the free streaming fixed point depends on higher moments through the value of the moment $\L_2$, as we have already observed. It follows that the attractor solution depends on $\L_2$. Elaborating on the discussions presented in \cite{Blaizot:2017ucy,Blaizot:2019scw,Blaizot:2021cdv},  we examine now various approximate ways to handle $\L_2$, without having to solve the full hierarchy of moment equations. 

Let us start by noticing that while we do not know $\L_2$ in general, we know its value in the vicinity of the free-streaming fixed points (whose locations are known exactly, namely $g_0=-1, -2$). To study the vicinity of the stable fixed point (which eventually turns into the hydrodynamic fixed point under the effects of collisions), we can then use the exact value of $\L_2$ near the stable fixed point, $\L_2=A_2 \L_0$ \cite{Blaizot:2019scw}. This has the effect of putting the stable fixed point at the right place, i.e., at  $g_0=-1$ (thereby eliminating the unphysical feature of a possible negative longitudinal pressure \cite{Blaizot:2019scw} that violates the bounds stated in Eq.~\eqref{bounds}). An alternative, and more accurate, treatment of the term $\L_2$ consists in noticing that \cite{Blaizot:2021cdv}
\begin{equation}\label{renorm_a1}
-c_0c_1 \frac{\L_2}{\L_0}=-c_1c_0 \frac{A_2}{A_1}\frac{\L_1}{\L_0}=c_1\frac{A_2}{A_1} (g_0+a_0),
\end{equation}
where  Eq.~(\ref{eq:g0def}) has been used. Referring back to Eq.~(\ref{eq:betafct}), We then observe that the term $c_1\frac{A_2}{A_1}$ can be absorbed into a redefinition of $a_1$, 
\begin{equation}\label{eq:a1prime}
a_1\mapsto \bar a_1=a_1+c_1\frac{A_2}{A_1}=\frac{31}{15}.
\end{equation}
In the massless case, this second strategy is the preferred one because it does not affect the behavior of the solution near the hydrodynamic fixed point. Recall that in the massless case, $\L_2/\L_0\sim 1/\tau^2$ at late time, which entails that $\L_2$ does not contribute to the first-order transport coefficient. In the first strategy, the constant term $c_0 c_1 A_2$ leaves an imprint at late time, effectively correcting the leading order viscosity coefficient ($b_1 c_0\mapsto b_1 c_0 +c_0 c_1 A_2$). That is, it  modifies the approach to the hydrodynamic fixed point, which we want to avoid%
    \footnote{One may note that $\L_2/\L_0$ as given by Eq.~(\ref{renorm_a1}) decays only as $1/\tau$ at late time, rather than $1/\tau^2$. However, in practice this is of no consequence since this behavior does not affect the leading-order viscosity.}.  
We shall follow a similar approach in the massive case, with a slight modification needed to take into account the non vanishing of the equilibrium moments in the vicinity of the hydrodynamic fixed point. 

In order to study how this adjustment of $\L_2$ modifies the attractor solution, we plot in Fig.~\ref{fig:g0_adiabatic}(b) the approximate solutions obtained by solving the equation $\beta(g_0(w),w)=0$, together with Eq.~(\ref{eq:zevol_g0}), with and without the adjustments of $\L_2$, and compare both choices just discussed. To avoid the precise determination of the last contribution in Eq.~(\ref{eq:betafct}) we use the approximate equation of state $P/\varepsilon=c_s^2$ and treat  $c_s^2$ as a constant. As anticipated the free-streaming fixed point is now well reproduced, while the hydrodynamic fixed point is not modified. The two adjustments of $\L_2$ yield very similar results. They differ only, slightly, in the transition region between free streaming and hydrodynamics, that takes place when $\tau\sim \tau_R$, i.e., when the expansion rate is comparable to the collision rate. The speed of sound also presents  a transition region from the massless regime to a non-relativistic regime, which occurs when $m\gtrsim T$ (see Fig.~\ref{fig:cs2_gstar}(b) and the discussion after Eq.~(\ref{eq:hydrog*})). Since we want to consider situations where the transition from the collisionless regime to the collision dominated regime takes place well before the systems enter the nonrelativistic regime, we shall focus on cases where the ratio $m/T$ remains small for times $\tau\sim \tau_R$, i.e. $m/T(\tau_R)\lesssim 1$ (a larger value, $m/T(\tau_R)=5$, will also be considered in order to exemplify the effect of a large mass). 

\begin{figure}[t!]
    \centering
    \includegraphics[width=\linewidth]{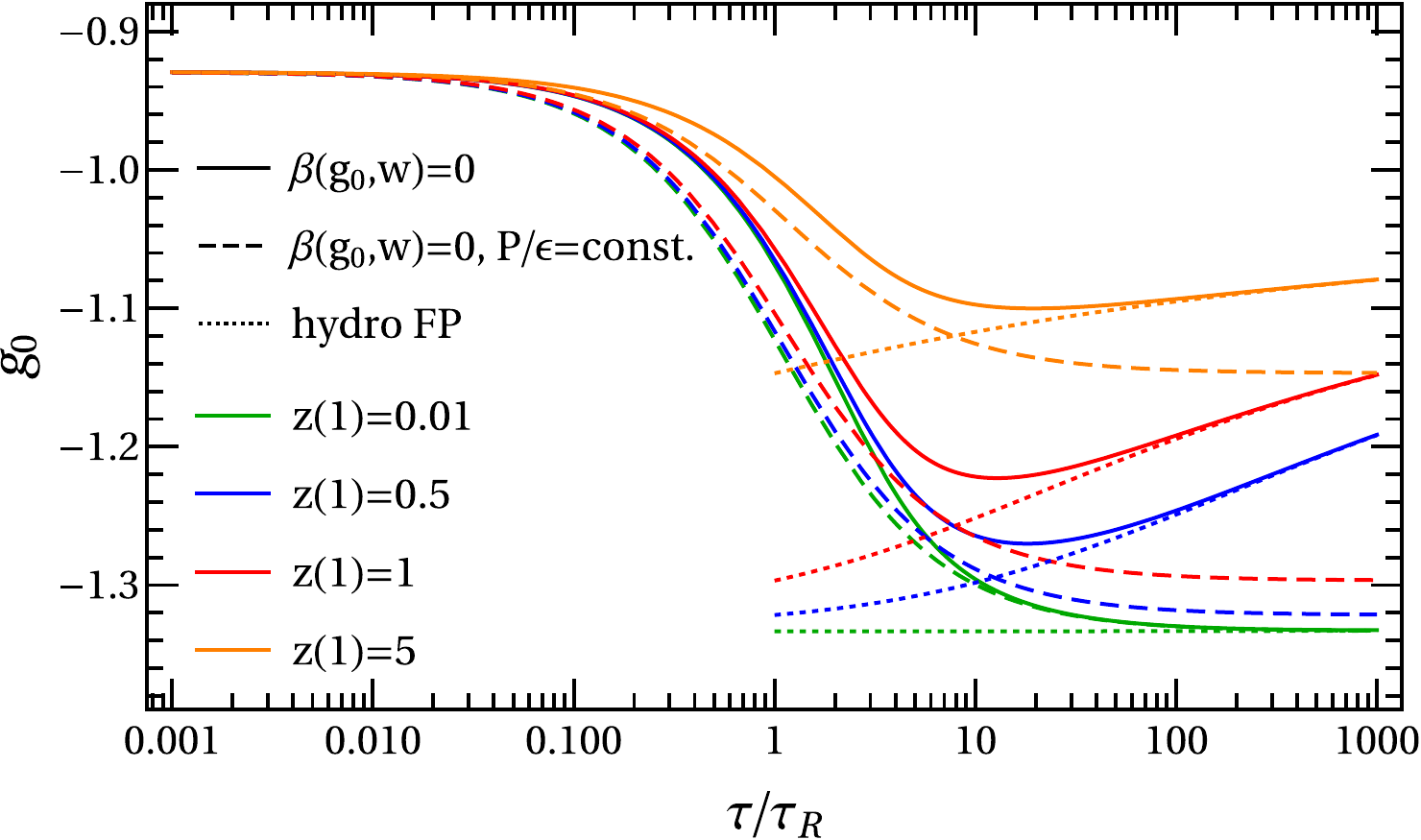}
    \vspace*{-6mm}
    \caption{The full curves represent the approximate attractor solution $g_0$ obtained by solving the implicit equation $\beta(g_0,w)=0$ together with Eq.~(\ref{eq:zevol_g0}). The dashed curves are obtained by freezing $P/\varepsilon$ at its value for $w=1$. The dotted curves represent the hydrodynamic fixed point, Eq.~\eqref{eq:g*_fp}. The green, blue, red, and orange curves, {\it in this figure and all figures after this}, correspond to $z(1)=m/T(\tau_R)=0.01,\, 0.5,\, 1,$ and $5$, respectively.}
    \vspace*{-2mm}
    \label{fig:g0_massive}
\end{figure}

We show in Fig.~\ref{fig:g0_massive} the approximate attractor solutions corresponding to values of $z(1)=m/T(\tau_R)$ ranging from $0.01$ to $5$ (solid lines). The characteristic drop of the attractor  for $\tau\gtrsim \tau_R$, which marks the transition to the collision-dominated regime and hydrodynamics, remains clearly visible in all cases. However, the solutions eventually bend upward at large times, reflecting the drop in the speed of sound (see Fig.~\ref{fig:cs2_gstar}(a)) as the temperature drops and the nonrelativistic regime is approached. The attractor then evolves in time according to Eq.~(\ref{eq:hydrog*}), that is, according to the evolution of the speed of sound (dotted lines). In order to contrast this behavior with that expected if the hydrodynamic fixed point was independent of time, we have repeated the calculation by freezing the ratio $P/\varepsilon$ at the value it takes for $\tau=\tau_R$; the corresponding solutions are shown as dashed curves. At late time, they reach  the fixed points  given by  Eq.~(\ref{eq:hydrog*}) for the corresponding constant values of the speed of sound (see Eq.~(\ref{eq:g*_fp})).

\vspace{-.3cm}
\subsubsection{Exact attractor solution}
\vspace{-.3cm}

\begin{figure*}[t!]
    \centering
    \includegraphics[width=\textwidth]{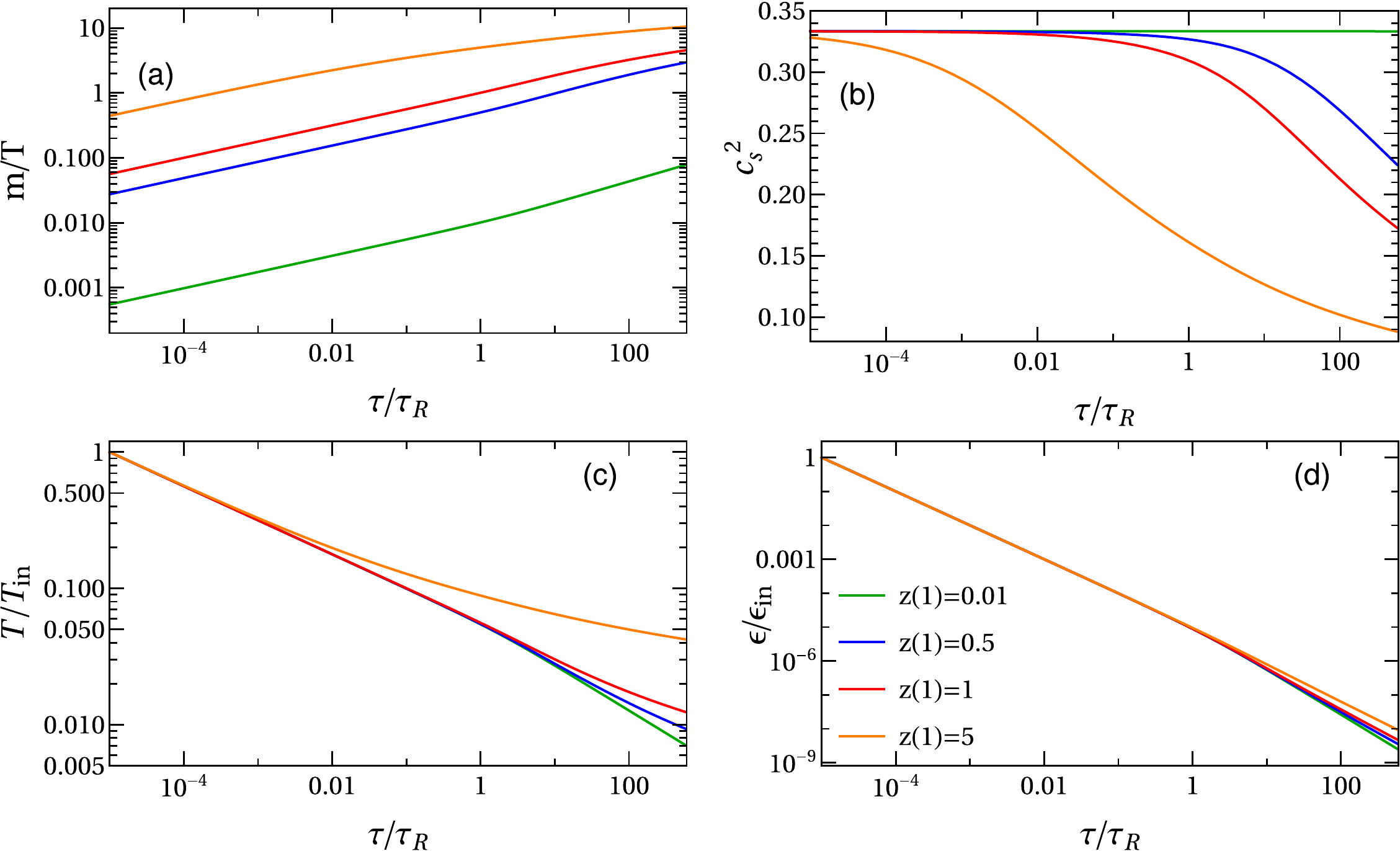}
    \vspace*{-6mm}
    \caption{Exact kinetic evolution of (a) $m/T$, (b) $c_s^2$, (c) $T/T_{\rm in}$, and (d) $\varepsilon/\varepsilon_{\rm in}$, for attractor initial condition with the following values of $z(1)=m/T(\tau_R):\;0.01, 0.5, 1,$ and $5$.}
    \label{fig:4new}
\end{figure*}

We turn now to a comparison with the exact solution of the kinetic equation. The calculations in this section  have been performed for values of $z(1)=m/T(\tau_R)$ ranging from $0.01$ to $5$, and initialized at time $\tau_{\rm in}=10^{-5}\tau_R$ with a  distribution peaked around $p_z=0$, corresponding to an initial condition near the stable free streaming fixed point (details of the numerical implementation can be found in Appendix~\ref{app:exactKT_sol}). The resulting time evolutions of $m/T$, of the ratio of the temperature to its initial value $T/T_{\rm in}$, and of the speed of sound squared $c_s^2$, are displayed in Fig.~\ref{fig:4new}. The increase of $m/T$ as a function of time (Fig.~\ref{fig:4new}(a)) follows from the decrease of the temperature (Fig.~\ref{fig:4new}(c)). The drop in the speed of sound, that accompanies the transition to the nonrelativistic regime, occurs for $\tau\gtrsim \tau_R$ when $z(1)\lesssim 1$, but is already significant at $\tau=\tau_R$ for $z(1)=5$. The energy density decreases as in free-streaming for $\tau \ll \tR$, i.e. $\varepsilon\sim 1/\tau$ (Fig.~\ref{fig:4new}(d)). Only for $\tau\gtrsim \tau_R$ is the behavior affected by the mass, via the dependence on the speed of sound. 

\begin{figure}[t!]
    \centering
	\includegraphics[width=\linewidth]{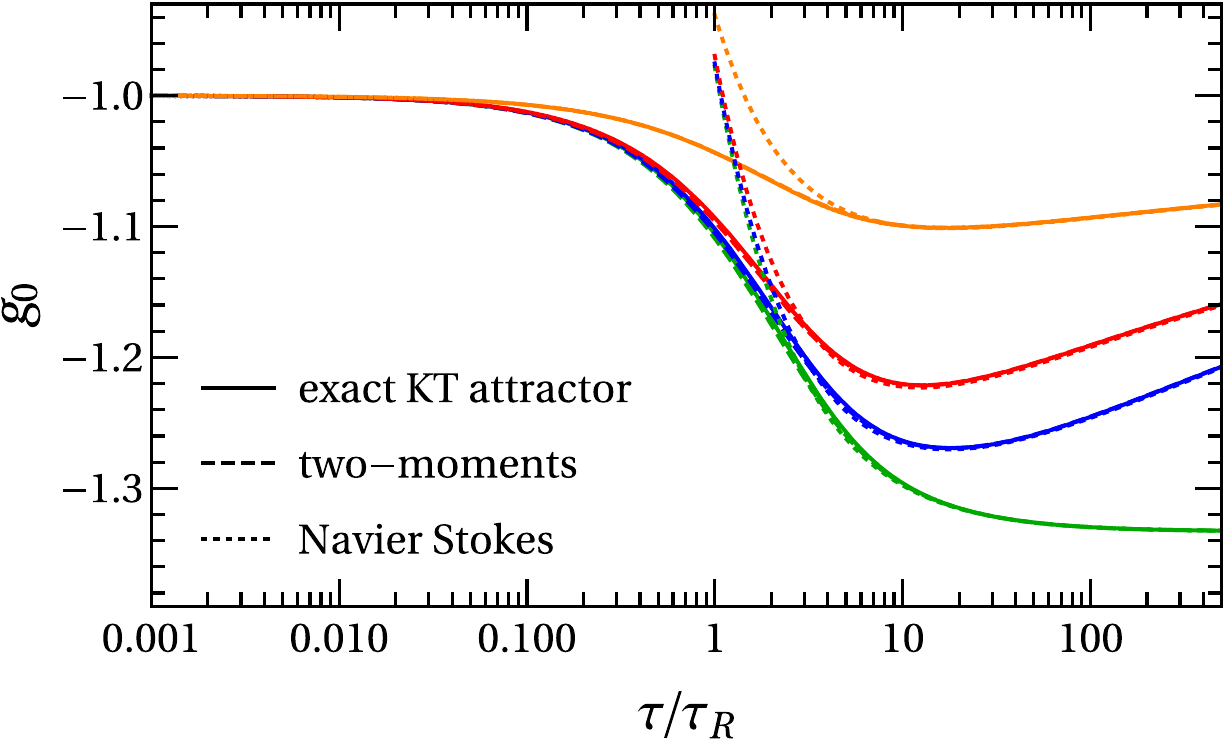}
	\vspace*{-6mm}
    \caption{The attractor solution for $g_0$: Comparison between the exact solution of the kinetic equation (full lines) and that of the two-moment truncation (dashed lines), with $\L_2$ given by Eq.~\eqref{ansatzL2}. The Navier-Stokes approximation (dotted lines) is obtained from the gradient expansion of $g_0$, $g_0(w)=\alpha_0+\alpha_1/w$, with the coefficient $\alpha_1$ given in Eq.~(\ref{3_16}).} 
    \vspace*{-2mm}
    \label{fig:g0_exact}
\end{figure}

The corresponding attractor solutions obtained from solving the kinetic equations are plotted in Fig.~\ref{fig:g0_exact} (solid curves). We compare them with the approximate attractors obtained from the truncated moment equations, using for $\L_2$ a slight generalization of Eq.~(\ref{renorm_a1}) that will be described shortly. The dashed lines that correspond to these approximate attractors are indistinguishable from the full lines. Also indicated in the figure are the results obtained with the Navier-Stokes hydrodynamics. As can be seen, this approximation reproduces well the attractor solution for $\tau\gtrsim \tau_R$. We recall now how this approximation is obtained.   

The collision terms in all the moment equations,  except that for $\L_0$, involve the deviations of the moments from their equilibrium values. These deviations are the quantities that are expected to have a gradient expansion, rather than the moments themselves (in the massless case, since the equilibrium values of the moment vanish, the moments themselves have a gradient expansion). The hydrodynamic fixed point has been already determined, and can be obtained by solving Eq.~(\ref{eq:betafct}). We can go one step further and use the gradient expansion to quantify the approach to the fixed point, which is controlled by Navier-Stokes hydrodynamics. To do so, we consider first the equation for $\L_1$ and set $\L_1'=\L_1 - \L_1^{\rm eq}$. By keeping the leading terms at large time we obtain, for $\L_1'$,
\begin{equation}\label{L1prime}
\L_1'=-\frac{\tau_R}{\tau} \left(b_1\L_0+a_1\L_1^{\rm eq}+ c_1\L_2^{\rm eq} \right)-\tau_R \frac{\rmd \L_1^{\rm eq}}{\rmd \tau}.
\end{equation}
A similar analysis performed for $\M_0$ yields
\begin{equation}\label{M0prime}
\M_0'=-\frac{\tau_R}{\tau} \left(a_0' \M_0^{\rm eq}+ c_0'\M_1^{\rm eq} \right)-\tau_R \frac{\rmd \M_0^{\rm eq}}{\rmd \tau}.
\end{equation}
As we shall see in the next section these  equations determine the shear and bulk viscosities, in agreement with the leading order of the Chapman-Enskog expansion%
    \footnote{The last terms in Eqs.~(\ref{L1prime}) and (\ref{M0prime}), that involve time derivatives of equilibrium quantities, actually contain second-order contributions. See Sec.~\ref{sec:newcoeff} and Appendix~\ref{app:gr}.}.

It is interesting to see how these viscosities emerge from Eq.~(\ref{eq:betafct}).
Assuming that $g_0(w)$ has a gradient expansion, we set $g_0(w)=\alpha_0+\alpha_1/w$, and solve  Eq.~(\ref{eq:betafct}) to leading order. We get 
\begin{align}
\alpha_0 \,=\,& g_*,
\nn
\alpha_1 \,=\,& b_1 c_0+c_1 c_0 \frac{\L_2^{\rm eq}}{\L_0}-a_0 a_1-(a_0+a_1)g_* -g_*^2
\nn
&+\frac{3c_0}{2}\left( c_s^2-\frac{P}{\varepsilon}  \right)g_*,
\end{align}
where (recall Eq.~(\ref{eq:g0def})) 
\begin{equation}
g_*=-a_0-c_0 \frac{\L_1^{\rm eq}}{\L_0}.
\end{equation}
The last term in the expression above for $\alpha_1$ originates from the variation of $\alpha_0=g*$ in the left-hand side of Eq.~(\ref{eq:betafct}). We have indeed
\begin{equation}
w\frac{{\rm d } g_*}{{\rmd }w}=-\frac{3c_0}{2}\left( c_s^2-\frac{P}{\varepsilon} \right) g_*.
\end{equation}
A simple calculation then yields
\begin{align}\label{3_16}
\alpha_1 &= b_1 c_0+c_0a_1\frac{\L_1^{\rm eq}}{\L_0}+c_1 c_0 \frac{\L_2^{\rm eq}}{\L_0}-\frac{c_0}{2} (1-3 c_s^2) g_*
\nn
&=-c_0 w \frac{\L_1'}{\L_0},
\end{align}
where $\L_1^{\rm eq}=-(\varepsilon-3P)/2$ has been used. One recovers the result obtained earlier for $\L_1'$, Eq.~(\ref{L1prime}), in particular, near the hydrodynamic fixed point, $g_0(w)$ takes the form
\begin{equation}\label{3_12}
g_0(w)=-a_0-c_0\frac{\L_1}{\L_0}=g_*-c_0\frac{\L_1'}{\L_0}.
\end{equation}

This analysis suggests that, in the massive case, in order to get a good approximation of $g_0(w)$ in the vicinity of the two fixed points, the stable free-streaming fixed point and the hydrodynamic fixed point, we need to fix $\L_2$ differently in either vicinity of the two fixed points. The following simple interpolation provides an excellent approximation to the exact solution: 
\begin{equation}\label{ansatzL2}
\L_2 =  \frac{{A}_2}{{A}_1} \L_1 + \mathcal{F}(w) \left(\L_2^{\rm eq}- \frac{{A}_2}{{A}_1}  \L_1^{\rm eq} \right) ,
\end{equation}
where  $\mathcal{F}(w)=1-e^{-w/2}$. In the free-streaming regime, $\mathcal{F}(w)\to 0$  and $\L_2 \to  \frac{{A}_2}{{A}_1} \L_1$: the free-streaming fixed point is correctly reproduced. For large $w$,  $\mathcal{F}(w) \to 1$, and $\L_1\to \L_1^{\rm eq}$, so that $\L_2\to \L_2^{\rm eq}$, as it should be near the hydrodynamic fixed point. Although \eqref{ansatzL2}   provides only a rough approximation to $\L_2$ in the transition region between the two fixed points, this is enough to obtain an excellent approximation to the full solution for the attractor solution, as can be seen in Fig.~\ref{fig:g0_exact} and will be documented further later in this section.

\subsection{The equation for the  moment \texorpdfstring{$\M_0$}{}}
\label{sec_FP_mn}
\vspace*{-2mm}

In order to solve Eq.~(\ref{eq_M0}), one may proceed again by analysing the vicinity of the two fixed points of Eq.~(\ref{eq:betafct}). Consider first the stable collisionless fixed point. We know that in its vicinity, the $\M$-moments are, like the $\L$-moments, proportional to each other, viz. $\L_n=A_n \,\L_0, \ \M_n=A_n\, \M_0$, where $A_n=P_{2n}(0)$ ($A_1=-1/2$, $A_2=3/8$). Consider then the evolution equation of the moment $\M_0$
\begin{equation}\label{eq:equationM0}
    \pder{\M_0}{w} = -\frac{1}{w}  \left( a'_0 \M_0 + c_0' \M_1 \right) - \left( \M_0 - \M_0^{\rm eq} \right)  \,.
\end{equation}
Near the stable free-streaming fixed point, as we just argued,  $\M_1/\M_0\simeq\mathcal{A}_1=-1/2$, and the last term $\propto (\M_0 - \M_0^{\rm eq}) $ can be ignored when $\tau\ll\tau_R$. We get then 
\begin{equation}
\pder{\M_0}{w} \simeq -\frac{\M_0}{w} \left( a'_0 + c_0' \frac{\M_1}{\M_0} \right) = - \frac{\M_0}{w} \,.
\end{equation}
Thus, in the vicinity of the stable free-streaming fixed point the effect of the moment $\M_1$ can be accounted for by a simple renormalization of the coefficient $a_o'$, i.e.,  $a_0' \mapsto  (a'_0  + c_0' \M_1/\M_0) = (a'_0  + c_0' A_1) =1$.

\begin{figure*}[t!]
    \centering
    \includegraphics[width=.95\textwidth]{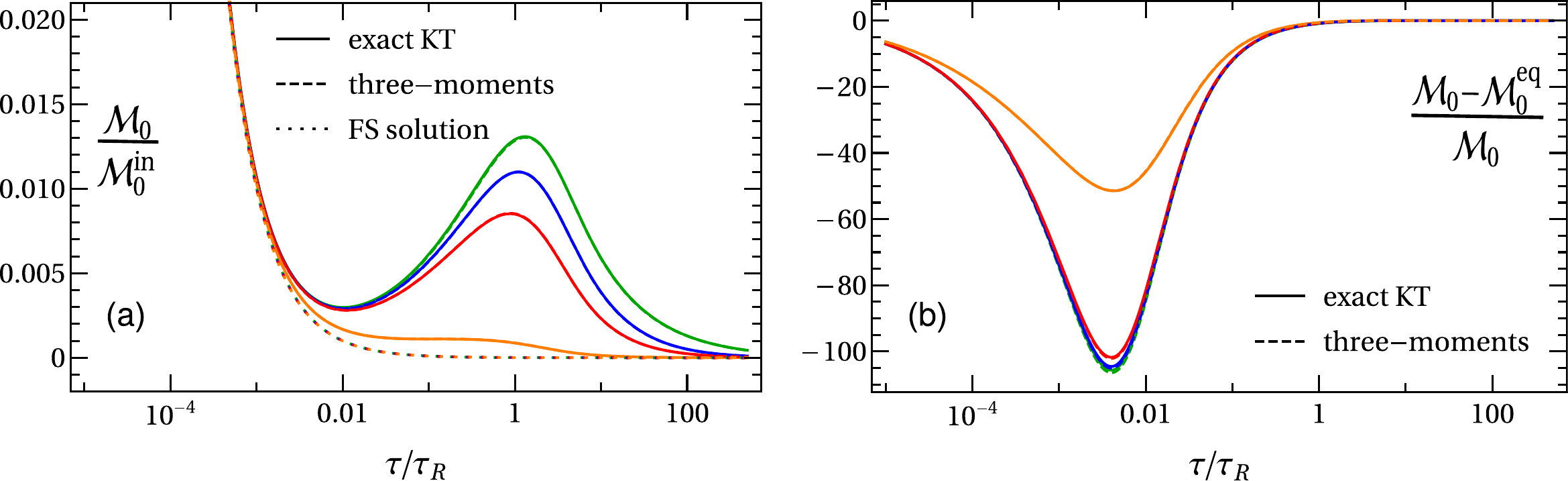}
    \vspace*{-.2cm}
    \caption{Evolution of (a) the moment $\M_0$ normalized with the initial value of $\M_0$, and (b) the relative deviation $(\M_0-\M_0^{\rm eq})/\M_0$. Note that the curves from bottom to top (orange, red, blue, and green) in (a) correspond to $z(1)=5,1,0.5$, and 0.01, respectively, while in (b), these appear in opposite order. The dotted line in panel (a) corresponds to the free-streaming (FS) evolution from the same initial condition. The results of the three-moment truncation are hardly distinguishable from the exact result.}
    \label{fig:M0_h0}
\end{figure*}

On the other hand, near the hydrodynamic fixed point, $\M_1\sim \M_1^{\rm eq}$. We can then, as we did earlier for $\L_2$, use a simple interpolation formula to connect the two fixed points. The formula 
\begin{equation}\label{ansatzM1}
\M_1 =  {A}_1 \M_0 +\mathcal{F}(w) \left( \M_1^{\rm eq}-A_1\M_0^{\rm eq} \right),
\end{equation}
where $\mathcal{F}(w)=1-e^{-w/2}$, analogous to (\ref{ansatzL2}) leads to an excellent agreement between the exact and the approximate $\M_0$. This is illustrated in Fig.~\ref{fig:M0_h0}(a), where differences between the exact solution and that of the three-moment truncation are hardly visible.

A noticeable feature of the plots in Fig.~\ref{fig:M0_h0} is the fact that collisions  appear to play a role at times that are much smaller than the collision time $\tau_R$: the deviation from the early time free-streaming behavior takes place when $\tau\lesssim 0.01 \tau_R$. To understand why this is so, we may consider the plot in Fig.~\ref{fig:M0_h0}(b). This plots shows that at early times, the rapid decay of $\M_0$ due to free streaming, makes this moment eventually much smaller than its equilibrium value $\M_0^{\rm eq}$. When  $|\M_0-\M_0^{\rm eq}|$ reaches its maximum, the effective expansion rate $\sim \M_0/\tau$ is of the same order of magnitude as the collision rate $\sim (\M_0^{\rm eq}- \M_0)/\tau_R$. From that point on the collisions take over and  force $\M_0$ to relax towards $\M_0^{\rm eq}$, a situation reached when $\tau\simeq \tau_R$ as can be seen in Fig.~\ref{fig:M0_h0}(b). Further decrease of $\M_0$ passed its maximum near $\tau_R$ is correlated to that of $\M_0^{\rm eq}$ with time. 

\begin{figure*}[t!]
    \centering
    \includegraphics[width=.95\textwidth]{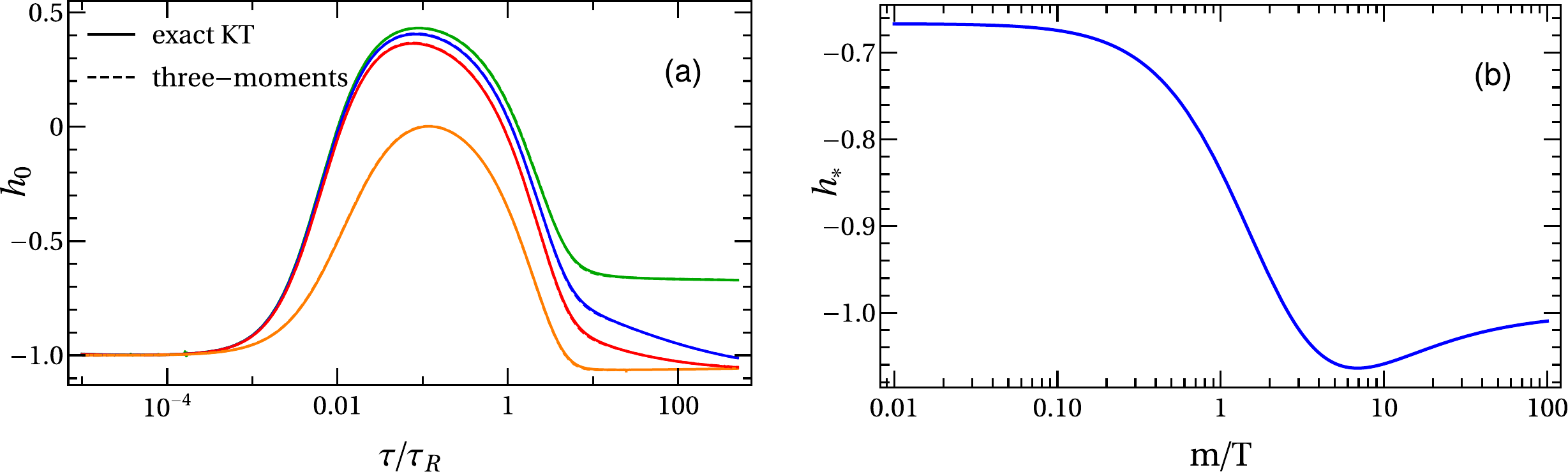}
    \vspace*{-.2cm}
    \caption{(a) Comparison of evolution of $h_0$ for exact and renormalized equations for $\M_0$. The curves from bottom to top (orange, red, blue, and green) correspond to $z(1)=5,1,0.5$, and 0.01, respectively. (b) the evolution of $h_*$  as a function of $m/T$.}
    \vspace*{-3mm}
    \label{fig:h0_h*}
\end{figure*}

The plot of the logarithmic derivative of $\M_0$,
\begin{equation}
h_0=\frac{w}{\M_0}\frac{\partial \M_0}{\partial w}
\end{equation}
reveals another interesting structure (Fig.~\ref{fig:h0_h*}). The plot in the left panel carries essentially the same information as that of the left panel of Fig.~\ref{fig:M0_h0}: at small time, $h_0=-1$, as expected in the free-streaming regime. Then $h_0$ exhibits a maximum and at late time returns to a value close to $-1$. Note that in the latter case, even though the power law of the decay is the same as in the collisionless regime, the physics is different. The situation is similar to what happens to $g_*$ for large values of $m/T$, as illustrated in Fig.~\ref{fig:cs2_gstar}(b). As we have already emphasized, the regime of large $m/T$, necessarily reached at late time if the mass is constant while the temperature decreases, is a nonrelativistic regime where the pressure becomes vanishingly small ($P/\varepsilon \sim 1/z \to 0$), and the particles follow the expansion without any intrinsic motion%
    \footnote{This non-relativistic regime has been referred to recently as an ``IR convergent point'' in Ref.~\cite{Kamata:2022jrc} which focuses on constructing trans-series solutions around this particular point.}.

In fact, values smaller than $-1$ are reached in the hydrodynamic regime. In this regime, 
\begin{align}
    h_0 \rightarrow h_*= &\frac{w}{\M_0^{\rm eq}}\frac{\partial \M_0^{\rm eq}}{\partial w} 
    = (1-3 c_s^2)\left(\frac{c_0}{2}-a_0\frac{\L_0}{\M_0^{\rm eq}} \right) 
    \nn
    =& - \left(\frac{\varepsilon+P}{\varepsilon-3P}\right) (1-3c_s^2) .
\end{align}
The evolution of $h_*$ as a function of $m/T$ is shown in Fig.~\ref{fig:h0_h*}(b). In the massless limit, we can use the small mass expansion for $1-3c_s^2=z^2/12+\mathcal{O}(z^3)$ and that for $1-3P/\varepsilon = z^2/6+\mathcal{O}(z^3)$ to verify that 
\begin{equation}
\lim_{z\to 0} h_* = -\left(\frac{1+P/\varepsilon}{1-3P/\varepsilon}\right) (1-3c_s^2) = -\frac{(4/3)}{(z^2/6)} \frac{z^2}{12} = -\frac{2}{3}.
\end{equation}
For large values of $m/T$, one can see from Fig.~\ref{fig:h0_h*}(b) that $h_*$ goes below $-1$ for $z\gtrsim 2$ and approaches $-1$ as $z\to \infty$. The latter case corresponds to very large masses and the non-relativistic regime that we have just discussed, where the pressure and the speed of sound vanish. This behavior of $h^*$ as a function of $z=m/T$ explains the behavior of $h_0$ seen in Fig.~\ref{fig:h0_h*}(a) at late times, after taking into account that $z$ increases as a function of time.

\begin{figure*}[t!]
    \centering
    \includegraphics[width=\textwidth]{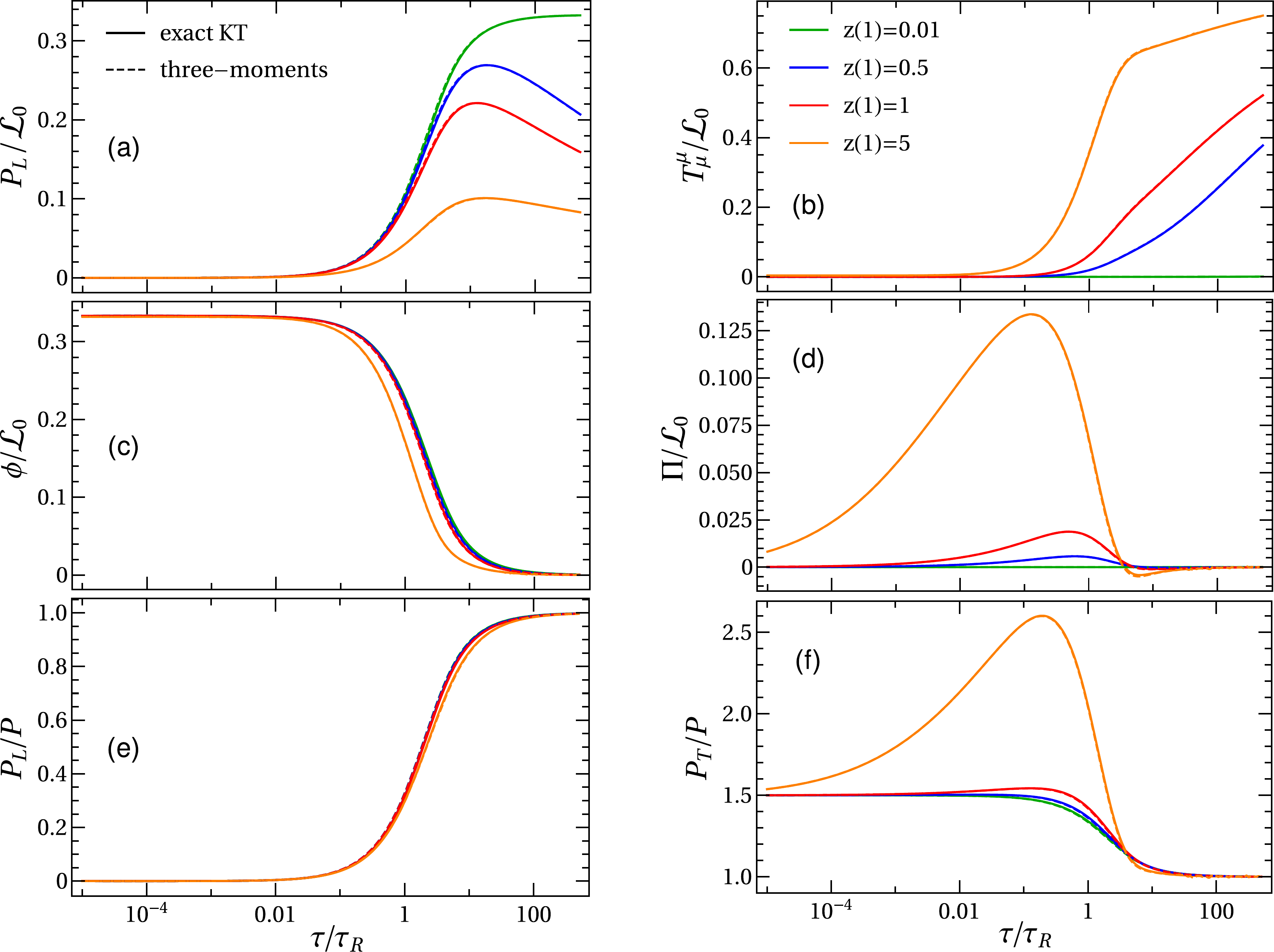}
    \vspace*{-.6cm}
    \caption{Comparison of solutions (dashed lines) obtained with the three-moment truncation (with the interpolation~(\ref{ansatzL2}, \ref{ansatzM1}) and the exact solutions of the full kinetic equation (solid lines). The initial time is $\tau_{\rm in}=10^{-5}\, \tau_R$. In panels (a), (c) and (e), the curves from bottom to top (orange, red, blue, and green) correspond to $z(1)=5,1,0.5$, and 0.01, respectively. In other panels, the color code is the same, but the curves appear ordered in the opposite way.}
    \vspace*{-.2cm}
    \label{fig_IC30}
\end{figure*}

\subsection{Summary of results for the three-moment truncation}
\vspace*{-2mm}

In this subsection, we present a set of results that illustrate the quality of the three-moment truncation as compared to the exact solution. We start with attractor solutions, namely solutions that evolve from the free-streaming fixed point at a small initial time. Then we consider isotropic initial conditions as an example of more general initial conditions.

\vspace*{-5mm}
\subsubsection{Attractor solutions} 
\vspace*{-3mm}

\begin{figure*}[t!]
    \centering
    \includegraphics[width=.95\textwidth]{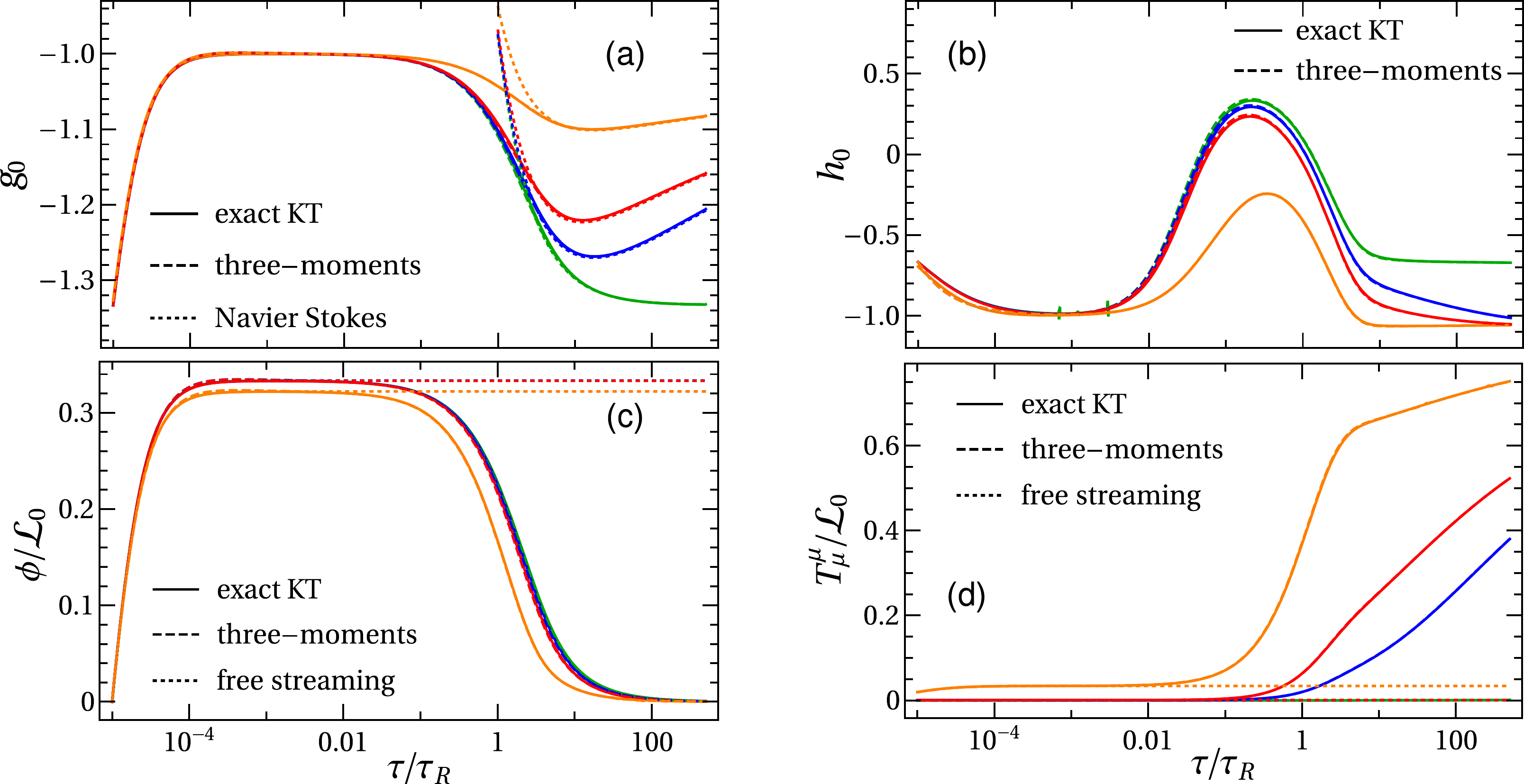}
    \vspace*{-.2cm}
    \caption{Comparison of solutions obtained with three-moment truncation (dashed lines) with exact kinetic theory (solid lines). The initial distribution is isotopic at $w_{\rm in}=10^{-5}$. The orange, red, blue, and green curves correspond to $z(1)=5,1,0.5$, and $0.01$, respectively. The dotted lines in the lower panels correspond to the free-streaming solution starting at the same initial conditions.}
    \label{fig:IC40}
\vspace*{-.2cm}
\end{figure*}

Attractor solutions are shown in  Figs.~\ref{fig_IC30}. As can be seen, the agreement of the three-moment truncation with the full kinetic theory is excellent for all quantities considered. The initial free-streaming regime is clearly visible in the plots of $P_L/\L_0$, $\M_0/\L_0$ and $\phi/\L_0$  (panels (a), (b), (c)). The interpretation of panel (a) representing $P_L/\L_0$ follows that of the attractor solution discussed earlier in this section (see Fig.~\ref{fig:g0_exact}). The two quantities are indeed simply related: from Eq.~(\ref{eq:g0def}) one deduces that $P_L/\L_0=-(g_0+1)$. The isotropization of the pressure at late time is confirmed in the plots of $P_L/P$, $P_T/P$ and also by that of $\phi/\L_0=(2/3) (P_T-P_L)/\L_0$ (panels (e), (f) and (c)).  The peculiar behaviors that can be seen in panels (d) and (f) representing respectively the ratio of the bulk pressure to the equilibrium pressure $\Pi/P$ and the equivalent ratio $P_T/P$ for the transverse pressure, have the following origin. From  the expression  (\ref{eq_isoP_phi}) for $P+\Pi$ one deduces
\begin{equation}
\frac{P}{\L_0}+\frac{\Pi}{\L_0}=\frac{1}{3}\left( 1-\frac{\M_0}{\L_0}\right).
\end{equation}
Near the free-streaming fixed point, both $\M_0$ and $\L_0$ behave in the same way (as $1/\tau$, see Sect.~\ref{sec:FS_regime}), that is, $\M_0/\L_0$ is a constant. Since $\M_0\ll \L_0$ when the mass is not too large, this ratio is very small at small time, as shown in panel (b). This implies that as long as $\M_0/\L_0\simeq 0$, the bulk pressure is essentially determined by the equilibrium pressure, whose evolution is tied to that of the effective temperature (via the Landau matching condition), i.e., $\Pi/\L_0\simeq 1/3-P/\varepsilon\simeq 1/3-c_s^2$. The variation of $c_s^2$ with time is shown in Fig.~\ref{fig:4new}, and it explains the rapid rise of $\Pi/\L_0$ that can be observed in panel (d) for small times and the largest value of the mass ($z=5$). Note that this feature occurs at times when the dynamics are not yet describable by hydrodynamics. Indeed, within the present sign convention, we would anticipate $\Pi$ to become negative once the hydrodynamic regime is reached, which is indeed the case at late time. However, when $\tau\lesssim \tau_R$, $\Pi$ takes positive values, which signals that kinetic, non-hydrodynamical effects are playing an important role there. Note that the time at which this transition takes place is about the same at which the Navier-Stokes solution starts to deviate from the exact kinetic solution, as seen in Fig.~(\ref{fig:g0_exact}). A similar discussion can be made for the transverse pressure. From Eq.~(\ref{transversepressure}), one deduces that 
\begin{equation}
\frac{P_T}{P}=\frac{P_T}{\L_0}\frac{\L_0}{P},\qquad \frac{P_T}{\L_0}=\frac{1}{3}\left(1-\frac{\L_1}{\L_0}-\frac{3}{2}\frac{\M_0}{\L_0}  \right).
\end{equation}
At small time, $P_T/\L_0$ is constant, so that the nonmonotonic behavior of $P_T/P$ observed in panel (f) for $z=5$ is to be attributed to that of the factor $\L_0/P$ which can be inferred from the plot in Fig.~\ref{fig:cs2_gstar}(a). 

\subsubsection{Isotropic initial conditions}
\vspace*{-3mm}

We consider now solutions more general than the attractor solutions, focusing for illustration purposes to the case of isotropic initial conditions. In this case, the interpolations that we used for attractor solutions need to be altered since the initial evolution is governed by free streaming starting away from the stable free-streaming fixed point. We then modify the interpolation formula (\ref{ansatzM1}) into
\begin{equation}\label{ansatzM1_iso}
\M_1 = \left(\! 1-\frac{w_{\rm in}}{w}\! \right)\! {A}_1 \M_0 +\mathcal{F}(w) \!\left[ \M_1^{\rm eq}- \! \left(\! 1-\frac{w_{\rm in}}{w} \!\right)\! A_1\M_0^{\rm eq} \right],
\end{equation}
which differs from  Eq.~\eqref{ansatzM1} solely by the term $\left( 1-\frac{w_{\rm in}}{w} \right)$ multiplying  the coefficient $A_1$. This term vanishes at initial time $w_{\rm in}$, making $\M_1^{\rm in}\to 0$ at the initial time. 
Similarly, we write the interpolation for $\L_2$ as 
\begin{equation}\label{ansatzL2_iso}
\L_2 = \left( 1-\frac{w_{\rm in}}{w} \right)\frac{{A}_2}{{A}_1} \L_1 + \mathcal{F}(w) \left[\L_2^{\rm eq}- \left( 1-\frac{w_{\rm in}}{w} \right)\frac{{A}_2}{{A}_1}  \L_1^{\rm eq} \right].
\end{equation}

In Fig.~\ref{fig:IC40}, we show the comparison of the exact solution of the kinetic equation  with that of the three-moment truncation with the above interpolations for $\M_1$ and $\L_2$, starting with an isotropic distribution at $w_{\rm in}=10^{-5}$. Again, the agreement between the exact solution and the three-moment truncation is excellent. In fact, because the initial time is small enough, the solution rapidly converges to the free-streaming fixed point, and then evolves along the attractor. We shall consider in the next section, a larger initial time where this does not happen, that is where the hydrodynamic fixed point starts to dominate the evolution of the system before it has time to reach the free-streaming fixed point  (see Fig.~\ref{fig:IC50}).

\vspace{-.2cm}
\section{Hydrodynamics}
\label{sec:mom_hydro}
\vspace{-.2cm}

We shall now compare the three moment equations to the second-order (Israel-Stewart) hydrodynamic equations, which we write in the form \cite{Denicol:2014vaa, Jaiswal:2014isa}:
\begin{subequations}
\label{hydro_evol}
\begin{align}
\label{eng_bj}
\frac{d\varepsilon}{d\tau} &= -\frac{1}{\tau}\left(\varepsilon + P + \Pi -\phi\right) \, ,  
\\ \label{bulk_bj}
\frac{d\Pi}{d\tau} + \frac{\Pi}{\tR} &= -\frac{\beta_\Pi}{\tau} - \delta_{\Pi\Pi}\frac{\Pi}{\tau}
+\lambda_{\Pi\phi}\frac{\phi}{\tau} \, ,  
\\ \label{shear_bj}
\frac{d\phi}{d\tau} + \frac{\phi}{\tR} &= \frac{4}{3}\frac{\beta_\phi}{\tau} - \left( \frac{1}{3}\tau_{\phi\phi}
+\delta_{\phi\phi}\right)\frac{\phi}{\tau} + \frac{2}{3}\lambda_{\phi\Pi}\frac{\Pi}{\tau}\,. 
\end{align}
\end{subequations}
Here $\beta_\phi \equiv \eta/\tR$ and $\beta_\Pi \equiv \zeta/\tR$ are the first-order transport coefficients related to the shear ($\eta$) and bulk ($\zeta$) viscosities. The second-order transport coefficients $\delta_{\Pi\Pi}, \,  \lambda_{\Pi\phi}, \,  \delta_{\phi\phi},\, \tau_{\phi\phi}$, and $\lambda_{\phi\Pi}$ are dimensionless functions of $m/T$ given in Ref. \cite{Jaiswal:2014isa}. The coefficients $\tau_{\phi\phi}$ and $\delta_{\phi\phi}$ have in general distinct origin, but because of the symmetry of Bjorken flow they only appear in the hydrodynamic equations in the combination indicated in Eq.~(\ref{shear_bj}).

\vspace{-.2cm}
\subsection{Derivation of hydrodynamic equations from the three-moment equations}
\label{sec:hydrofrommom}
\vspace{-.2cm}

There are strong similarities between the three-moment equations~(\ref{L_n:trunc}) and the hydrodynamic equations~\eqref{hydro_evol}. In fact we anticipate that these hydrodynamic equations describe accurately the late-time behaviour of the moment equations, which, as we have seen in the previous section, provide an almost exact solution of the kinetic equation~(\ref{keq_2}). To make the connection more precise, we shall show how the hydrodynamic equations can be derived from the moment equations. 

Consider first the equation for the energy density, Eq.~(\ref{eq_L_0}). By expressing the moments in terms of the hydrodynamic fields, using the relations 
$\L_0=\varepsilon$, $\L_1 =(1/c_0) (P_L-\varepsilon/3)$ $= -\phi/c_0-\M_0/2$, $\M_0=\varepsilon-3(P+\Pi)$, one obtains from Eq.~\eqref{eq_L_0},
\begin{equation}\label{eqforeps}
\pder{\varepsilon}{\tau} = -\frac{1}{\tau} \left(a_0\, \varepsilon +P_L -\varepsilon/3  \right) 
	= -\frac{1}{\tau} \left( \varepsilon +P +\Pi -\phi \right),
\end{equation}
which is the same as the energy evolution equation~\eqref{eng_bj}. This is to be expected since this equation just translates the energy-momentum conservation. Assuming that $\Pi$ and $\phi$ have a gradient expansion, such that $\Pi\sim 1/\tau,\, \phi\sim 1/\tau$ in leading order, we can drop these terms at late times. We then recover the ideal hydrodynamical evolution of the energy density that corresponds to the hydrodynamic fixed point (\ref{eq:hydrog*}). By keeping in the equation the leading-order contributions to $\Pi$ and $\phi$, one obtains the Navier-Stokes equation that controls the approach to the fixed point. These leading-order expressions of $\Pi$ and $\phi$ will be recalled shortly.

Consider now the equation for $\Pi$, which we shall obtain from Eq.~(\ref{eq_M0}). We need
\begin{equation}
\frac{\del (\varepsilon-3P)}{\del \tau} =(1-3c_s^2) \frac{\del \varepsilon}{\del \tau},\qquad c_s^2=\frac{\del P}{\del\varepsilon},
\end{equation}
and 
\begin{equation}
\Pi=-(\M_0-\M_0^{\rm eq})/3, \qquad \M_0^{\rm eq}=\varepsilon-3P.
\end{equation}
We then obtain from Eq.~(\ref{eq_M0})
\begin{equation}
\frac{d\Pi}{d\tau} + \frac{\Pi}{\tR} = \frac{1}{3\tau}\left(a_0'\M_0+c_0' \M_1\right)+\frac{1}{3}(1-3c_s^2)\frac{\del \varepsilon}{\del \tau}.
\end{equation}
We can go one step further and set $\M_0=\M_0^{\rm eq}-3\Pi $ and $\M_1=\M_1^{\rm eq}+\M_1'$, where $\M_1^{\rm eq}$ is a known quantity, but $\M_1'$ is unknown (in contrast to $\M_0' \equiv \M_0-\M_0^{\rm eq} \sim \Pi$). One may, however, assume that $\M_1'$, which measures the deviation from local equilibrium, has a gradient expansion with a leading order $\propto 1/\tau$. We then rewrite the equation above as
\begin{align}\label{bulkeqn_mom}
\frac{d\Pi}{d\tau} + \frac{\Pi}{\tR} =& \frac{1}{3\tau}\left(a_0'\M_0^{\rm eq}+c_0' \M_1^{\rm eq}\right)-a_0'\frac{\Pi}{\tau} + \frac{c_0'}{3} \frac{\M_1'}{\tau}
\nn
&+ \frac{1}{3}(1-3c_s^2)\frac{\del \varepsilon}{\del \tau},
\end{align}
where $\partial\varepsilon/\partial\tau$ can be expressed in terms of the hydrodynamic fields using Eq.~(\ref{eqforeps}). This equation is still exact, but it contains the unknown quantity $\M_1'$ (by unknown, we mean a quantity that cannot be expressed naturally in terms of the hydrodynamic fields). We shall deal with it at a later stage. At this point, one can separate the leading-order terms in $1/\tau$ and get
\begin{equation}\label{eq:PiLO}
\Pi=\frac{\tau_R}{3\tau}\left[ a_0'\M_0^{\rm eq}+c_0' \M_1^{\rm eq}-(1-3c_s^2)(\varepsilon+P) \right],
\end{equation}
where we have substituted in Eq.~(\ref{bulkeqn_mom}) $\partial \varepsilon/\partial \tau$ by its leading order contribution in Eq.~(\ref{eqforeps}).
The expression for the Navier-Stokes bulk viscosity follows
\begin{equation}
\zeta=\frac{\tau_R}{3}\left[ (1-3c_s^2)(\varepsilon+P)-\left(a_0'\M_0^{\rm eq}+c_0' \M_1^{\rm eq}\right) \right],
\end{equation}
where the quantity within the square brackets stands for $3\beta_\Pi$ ($\zeta=\tau_R \beta_\Pi$). Since no approximation has been done, beyond the leading-order gradient expansion, this should agree with the result obtained from the first-order Chapman-Enskog expansion, and indeed it does, as we verify in the Appendix~\ref{app:CE_exp}. 

To obtain the evolution equation for $\phi$ we note that 
\begin{equation}
\phi=-c_0\L_1-\frac{1}{3} \M_0,\quad c_0\L_1'\equiv c_0(\L_1-\L_1^{\rm eq})=\Pi-\phi,
\end{equation}
and we combine the equations (\ref{eq_L_1}) for $\L_1$ and (\ref{eq_M0}) for $\M_0$. We obtain
\begin{align}
\frac{d\phi}{d\tau}=& \frac{c_0}{\tau}\left( a_1\L_1+b_1\L_0+c_1\L_2 \right)+\frac{\Pi-\phi}{\tau_R}
\nn
&+\frac{1}{3\tau} \left(a_0'\M_0+c_0'\M_1 \right)+\frac{\M_0'}{3\tau_R}
\end{align}
or
\begin{align}
\frac{d\phi}{d\tau} + \frac{\phi}{\tR} = \frac{1}{\tau}\bigg[ &-a_1\phi-\frac{1}{3} a_1\M_0+c_0b_1\varepsilon+c_0c_1\L_2
\nn
&+\frac{1}{3}a_0'\M_0+\frac{1}{3}c_0'\M_1 \bigg].
\end{align}
Note the cancellation of the terms $\Pi/\tau_R$. At this point, we proceed as we just did for $\Pi$ and set  
 $\L_2= \L_2^{\rm eq}+\L_2'$. We then rewrite the above equation as 
\begin{align}\label{sheareqn_mom}
\frac{d\phi}{d\tau} + \frac{\phi}{\tR} =\frac{1}{\tau}\bigg[ &\left\{c_0b_1\varepsilon  +\frac{a_0'-a_1}{3}\M_0^{\rm eq}+\frac{c_0'}{3} \M_1^{\rm eq}+c_0c_1\L_2^{\rm eq}\right\} 
\nn
&- (a_0'-a_1)\Pi-a_1\phi + \frac{c_0'}{3} \M_1' +c_0c_1 \L_2' \bigg].
\end{align}
This equation is still exact, but it contains the unknown quantities, $\L_2'$ and $\M_1'$. It is nevertheless  easy  to isolate the leading-order terms of the gradient expansion: 
\begin{equation}\label{eq:PhiLO}
\phi=\frac{\tau_R}{\tau}\left[c_0 b_1\varepsilon +\frac{a_0'-a_1}{3}\M_0^{\rm eq}+\frac{c_0'}{3} \M_1^{\rm eq}+c_0c_1\L_2^{\rm eq}\right].
\end{equation}
The expression of the shear viscosity $\eta$ follows from the relation $\phi=2c_0\eta/\tau$.
It agrees with that obtained from the first-order Chapman-Enskog expansion (see Appendix~\ref{app:CE_exp}). 

\vspace{-.2cm}
\subsection{Second-order coefficients}
\vspace{-.2cm}

We turn now to the second-order coefficients. By comparing the second-order terms in the equations~(\ref{bulk_bj}) and (\ref{bulkeqn_mom}) for $\Pi$ (in which we substitute $\partial \varepsilon/\partial \tau$ by its second-order contribution in Eq.~(\ref{eqforeps}), we are led to the identification
\begin{equation}\label{coefdeltaPi}
\lambda_{\Pi\phi} \phi - \delta_{\Pi\Pi}\Pi = \left(\frac{1}{3} -c_s^2 \right) \phi -\left(a_0' +\frac{1}{3}-c_s^2\right) \Pi +\frac{c_0'}{3} \M_1'.
\end{equation}
In order to extract from this the values of the transport coefficients, we need to express $\M_1'$ in terms of $\Pi$ and $\phi$, which can be done using the Chapman-Enskog expansion. Note that only the leading-order expansion of $\M_1'$ is needed since it will be eventually multiplied by a gradient ($\sim 1/\tau$) in Eq.~(\ref{bulkeqn_mom}). With this in mind, the identification of the coefficients of $\Pi$ and $\phi$ in the above equation yields
\begin{align}\label{bulk_coeff}
\delta_{\Pi\Pi} &= a_0' +\frac{1}{3}-c_s^2 - \frac{c_0'}{3}  \Pi_{ \in \M_1'}  ,
\nn
\lambda_{\Pi\phi} &= \frac{1}{3}-c_s^2 +\frac{c_0'}{3} \phi_{ \in \M_1'},  
\end{align}
where the notation $\Pi_{ \in \M_1'}$ denotes the leading-order contribution of $\M_1'$ to $\Pi$, and similarly for $\phi_{ \in \M_1'}$. These contributions will be determined shortly. 

Similarly, by comparing the second-order terms in Eq.~\eqref{shear_bj} for the shear pressure $\phi$ with those of the exact evolution equation~\eqref{sheareqn_mom}, one gets
\begin{align}
- \left( \frac{1}{3}\tau_{\phi\phi} +\delta_{\phi\phi}\right) \phi + \frac{2}{3}\lambda_{\phi\Pi}\Pi = &-a_1\phi - (a_0'-a_1)\Pi 
\nn
&+ \frac{c_0'}{3} \M_1' +c_0c_1 \L_2',
\end{align}
from which, using the same notation as above, we obtain the following expressions of the transport coefficients
\begin{align}\label{shear_coeff}
 \frac{1}{3}\tau_{\phi\phi} +\delta_{\phi\phi} &= a_1 -\frac{c_0'}{3} \phi_{ \in \M_1'} -c_0c_1  \phi_{ \in \L_2'} ,
\nn
\lambda_{\phi\Pi} &= \frac{3}{2} (a_1-a_0') + \frac{c_0'}{2} \Pi_{ \in \M_1'} + \frac{3}{2}c_0c_1 \Pi_{ \in \L_2'}.
\end{align}

We turn now to the explicit calculation of  the leading-order contribution of $\M_1'$ and $\L_2'$ to $\Pi$ and $\phi$. As already indicated, this can be done using the Chapman-Enskog expansion (we refer to Appendix~\ref{app:CE_exp} for details). The leading-order correction to the equilibrium distribution function $f^{\rm eq}(p_0/T)$  is written as 
\begin{equation}\label{deltaf1_corr}
\delta f^{(1)}=\frac{\tau_R}{\tau}\frac{1}{p_0}\frac{\del f^{\rm eq}}{\del p_0}\left[ \left(\frac{p^2}{3}-c_s^2 p_0^2\right)+\frac{2}{3}q^2\right]=\mathcal{B}+\mathcal{S}q^2.
\end{equation}
Here $p^2=p_z^2+p_\perp^2$,  $q^2=p_z^2-p_\perp^2/2$, and $p_z^2=(p^2+2 q^2)/3$. In the second equality, we have separated the radial and angular deformation to the equilibrium distribution function, which we refer to as $\mathcal{B}$ and $\mathcal{S}$ respectively, 
\begin{equation}
\mathcal{B} \equiv  \frac{\tR}{\tau} \frac{1}{p_0} \left( \frac{p^2}{3} -c_s^2 p_0^2 \right) \pder{f^{\rm eq}}{p_0} , \quad 
\mathcal{S} \equiv \frac{2\tR}{3\tau} \frac{1}{p_0} \pder{f^{\rm eq}}{p_0} .
\end{equation}
A simple calculation then yields
\begin{align}\label{eq_M1p}
\M_1' =& \int_\p \frac{m^2}{p_0} P_2(p_z/p_0) \delta f^{(1)} = \left(\M_1'\right)_{\mathcal{B}} + \left(\M_1'\right)_{\mathcal{S}}
\nn
=& \left[\frac{\tR}{\tau} \left(\frac{m^6}{2} \int_\p \frac{1}{p_0^5} f^{\rm eq} + c_s^2 T \frac{d\M_1^{\rm eq}}{dT} \right) \right]_\mathcal{B} 
\nn
   &+\left[\frac{\tR}{\tau} \left(\frac{2m^6}{5} \int_\p \frac{1}{p_0^5} f^{\rm eq} - \frac{4}{15} (\M_0^{\rm eq}-\M_1^{\rm eq}) \right) \right]_\mathcal{S} ,
\end{align}
where we have separated the contributions coming from the terms ${\mathcal{B}}$ and ${\mathcal{S}}$ for a reason that we discuss now. 

In order to get the explicit values of the transport coefficients, it is convenient to eliminate the factor $\tau_R/\tau$ that is multiplying both contributions to $\M_1'$, in favor of the hydrodynamic fields $\Pi$ and $\phi$ using leading-order relations. There is obviously some ambiguity in doing so, to which we return in the next subsection. Here, we note that the factor $\tau_R/\tau$ appearing in $[\cdots]_\mathcal{B}$ in Eq.~\eqref{eq_M1p} arises from the radial deformation of the equilibrium distribution function and is proportional to $\partial_\mu u^\mu$, where $u^\mu$ is the collective four-velocity of the medium. On the other hand, the factor $\tau_R/\tau$ that appears in $[\cdots]_\mathcal{S}$ in Eq.~\eqref{eq_M1p} arises from the angular deformation and is proportional to $\partial^{\langle\mu}u^{\nu\rangle}$ where the angular brackets, $A^{\langle\mu\nu\rangle}$ represent the traceless symmetric projection of $A^{\mu\nu}$. It is therefore natural to use Eqs.~\eqref{tauRtau_B} and \eqref{tauRtau_S} for substituting $\tau_R/\tau$ in the first and second terms of the r.h.s of Eq.~\eqref{eq_M1p}, respectively. This is similar to the procedure followed in \cite{Jaiswal:2014isa}, and this is  the one that we  adopt in this subsection.  

Thus, in the first term of Eq.~(\ref{eq_M1p}), we replace $\tR/\tau$  by the leading-order (Navier-Stokes) expression for $\Pi$ taken from Eq.~\eqref{eq:PiLO}, i.e.,
\begin{equation}\label{tauRtau_B}
 \frac{\tau_R}{\tau} = \frac{3\Pi}{\left[ a_0'\M_0^{\rm eq}+c_0' \M_1^{\rm eq}-(1-3c_s^2)(\varepsilon+P) \right]}.
\end{equation}
A similar replacement for $\tR/\tau$ in the second term on r.h.s in Eq.~\eqref{eq_M1p} is done using the Navier-Stokes solution~\eqref{eq:PhiLO} for $\phi$,
\begin{equation}\label{tauRtau_S}
\frac{\tau_R}{\tau}= \frac{\phi}{\left[c_0b_1\varepsilon  +\frac{a_0'-a_1}{3}\M_0^{\rm eq}+\frac{c_0'}{3} \M_1^{\rm eq}+c_0c_1\L_2^{\rm eq}\right]}.
\end{equation}
\begin{widetext}
We then obtain the following expressions 
\begin{align}\label{PiinM1}
\Pi_{ \in \M_1' } &= \frac{3}{\left[ a_0'\M_0^{\rm eq}+c_0' \M_1^{\rm eq}-(1-3c_s^2)(\varepsilon+P) \right]} \left(\frac{m^6}{2} \int_\p \frac{1}{p_0^5} f^{\rm eq} + c_s^2 T \frac{d\M_1^{\rm eq}}{dT} \right) ,
\nn
\phi_{ \in \M_1'} &= \frac{1}{\left[c_0b_1\varepsilon  +\frac{a_0'-a_1}{3}\M_0^{\rm eq}+\frac{c_0'}{3} \M_1^{\rm eq}+c_0c_1\L_2^{\rm eq}\right]} \left(\frac{2m^6}{5} \int_\p \frac{1}{p_0^5} f^{\rm eq} - \frac{4}{15} (\M_0^{\rm eq}-\M_1^{\rm eq}) \right).
\end{align}

Proceeding in a similar fashion for $\L_2'$, one obtains first the  leading-order correction in the form 
\begin{align}\label{eq_L2p}
\L_2' &= \int_\p p_0\, P_4(p_z/p_0)\, \delta f^{(1)} = \left(\L_2'\right)_{\mathcal{B}} + \left(\L_2'\right)_{\mathcal{S}}
\nn
&= \left[\frac{\tR}{\tau} \left( c_s^2 T \frac{d\L_2^{\rm eq}}{dT} + \frac{(\M_0^{\rm eq}-\M_1^{\rm eq})}{3}-\frac{7 m^6}{8} \int_\p \frac{1}{p_0^5} f^{\rm eq} \right) \right]_\mathcal{B} +\left[\frac{\tR}{\tau} \left( \frac{2}{3} (\M_0^{\rm eq}-\M_1^{\rm eq}) -m^6 \int_\p \frac{1}{p_0^5} f^{\rm eq} \right) \right]_\mathcal{S} .
\end{align}

Substituting the factors $\tR/\tau$ by the appropriate factors (\ref{tauRtau_B}) or  (\ref{tauRtau_S}), as we did for $\M_1'$, we get
\begin{align}\label{PiinL2}
\Pi_{ \in \L_2'} &= \frac{3}{\left[ a_0'\M_0^{\rm eq}+c_0' \M_1^{\rm eq}-(1-3c_s^2)(\varepsilon+P) \right]} \left( c_s^2 T \frac{d\L_2^{\rm eq}}{dT} + \frac{(\M_0^{\rm eq}-\M_1^{\rm eq})}{3}-\frac{7 m^6}{8} \int_\p \frac{1}{p_0^5} f^{\rm eq} \right),
\nn
\phi_{ \in \L_2'} &= \frac{1}{\left[c_0b_1\varepsilon  +\frac{a_0'-a_1}{3}\M_0^{\rm eq}+\frac{c_0'}{3} \M_1^{\rm eq}+c_0c_1\L_2^{\rm eq}\right]}  \left( \frac{2}{3} (\M_0^{\rm eq}-\M_1^{\rm eq}) -m^6 \int_\p \frac{1}{p_0^5} f^{\rm eq} \right).
\end{align}
\end{widetext}

This completes the determination of the second-order transport coefficients. When the expressions (\ref{PiinM1}) and (\ref{PiinL2}) are plugged into the formulae~(\ref{bulk_coeff}, \ref{shear_coeff}), one obtains explicit values for these  coefficients. We have verified  that these values agree with those obtained using a similar method in Ref.~\cite{Jaiswal:2014isa}. Note that in order to derive these second-order coefficients, we needed only the first order Chapman-Enskog expansion (i.e., only $\delta f^{(1)}$). We verify in Appendix~\ref{app:gr}, by a direct solution of the moment equations, that these coefficients reproduce correctly the gradient expansion of the moments up to second order. 

\vspace{-.2cm}
\subsection{Ambiguity of second-order transport coefficients}
\label{sec:newcoeff}
\vspace{-.2cm}

We have observed in the previous section that the equations for $\L_0$ and $\L_1$ decouple from the equation for $\M_0$ (see the discussion after Eq.~(\ref{L_n:coeff_rel})). Since in the equation for the evolution of the energy density, Eq.~(\ref{eqforeps}), only $\Pi-\phi$ enters, and since $\L_1$ is given by 
\begin{equation}
\L_1=\L_1^{\rm eq}+\frac{1}{c_0}(\Pi-\phi), 
\end{equation}
one expects a similar decoupling of the hydrodynamic equations. However, if one rewrites the hydrodynamics equation~(\ref{bulk_bj}, \ref{shear_bj}) in the form
\begin{align}
\frac{d(\Pi-\phi)}{d\tau} +\frac{\Pi-\phi}{\tR} = -\frac{1}{\tau} \bigg[&\beta_\Pi+ \frac{\beta_\phi}{2} +\left( \delta_{\Pi\Pi}+\frac{2}{3}\lambda_{\phi\Pi}\right) \Pi
\nn
 &-\left( \lambda_{\Pi\phi} +\frac{1}{3}\tau_{\phi\phi}+\delta_{\phi\phi} \right)\phi \bigg],
\end{align}
the only way the above equation can be written completely in terms of  $\Pi-\phi$ is if
\begin{equation}\label{eq_condition}
\delta_{\Pi\Pi}+\frac{2}{3}\lambda_{\phi\Pi} =\lambda_{\Pi\phi} +\frac{1}{3}\tau_{\phi\phi}+\delta_{\phi\phi}, 
\end{equation}
which does not hold. 

To analyse the origin of this apparent discrepancy, let us return to the equation for $\L_1'$,
\begin{equation}\label{l1peq}
\frac{\del \L_1'}{\del \tau}+\frac{\L_1'}{\tau_R} =-\frac{1}{\tau}\left( a_1 \L_1'+ a_1 \L_1^{\rm eq}+ b_1\varepsilon+c_1\L_2+\tau\frac{d\L_1^{\rm eq}}{d\tau}\right).
\end{equation}
It is straightforward to verify that in leading order at large $\tau$, this equation reproduces the right combination of shear and bulk pressures compatible with the Chapman-Enskog expansion of $\Pi-\phi$. Beyond leading order, we need a proper treatment of the moment $\L_2$. By using Eq.~(\ref{eqLeqepsilon}), one can rewrite Eq.~\eqref{l1peq} as 
\begin{align}\label{eqforL1prime}
\frac{\del \L_1'}{\del \tau}+\frac{\L_1'}{\tau_R} =&-\frac{1}{\tau}\bigg[\bigg\{ a_1 \L_1^{\rm eq}+b_1\L_0+ c_1\L_2^{\rm eq}
\nn
& \qquad \quad -\left( a_0 + c_0 \frac{\L_1^{\rm eq}}{\L_0} \right) \pder{\L_1^{\rm eq}}{\varepsilon} \L_0 \bigg\}
\nn
&+\left(a_1-c_0 \pder{\L_1^{\rm eq}}{\varepsilon} \right) \L_1'+c_1\L_2'\bigg],
\nn
=& -\frac{1}{\tau}\left[s_1 \L_0 + (a_1+t_1)\L_1' + c_1 \L_2'  \right].
\end{align}
The coefficients $s_n, t_n, u_n, v_n$ are defined in Appendix~\ref{app:gr}. Both $\L_1'$ and $\L_2'$ are proportional to $\tau_R/\tau$. For $\L_1'$ we have indeed, in leading order of Chapman-Enskog expansion,
\begin{equation}\label{l1pns}
\L_1' = -\frac{\tR}{\tau} s_1 \L_0.
\end{equation}
As for $\L_2'$, it is given, at the same order by Eq.~(\ref{L2prime}), viz.
\begin{align}\label{L2prime_1}
\L_2' =& -\frac{\tau_R}{\tau} \left[ c_s^2 T \frac{\partial \L_2^{\rm eq}}{\partial T} + \M_0^{\rm eq}-\M_1^{\rm eq} -\frac{15m^6}{8} \int_\p \frac{f^{\rm eq}}{p_0^5} \right] \L_0 
\nn
=& -\frac{\tau_R}{\tau} s_2 \L_0.
\end{align}
This expression of $\L_2'$ differs from that used in the previous subsection, where the contributions of bulk and shear were separated. We shall see shortly that this leads to values for transport coefficients that differ from those obtained in Ref.~\cite{Jaiswal:2014isa}, while preserving the overall gradient expansion. 
Eq.~\eqref{l1pns} allows us to express the gradient $\tR/\tau$ in Eq.~(\ref{L2prime_1}) in terms of $\L_1'$ 
\begin{equation}\label{l2pasl1p}
\L_2' = \frac{s_2}{s_1} \L_1'.
\end{equation}
Substituting this expression in  Eq.~\eqref{eqforL1prime} one obtains an equation where only $\L_1'$ appears, 
\begin{equation}\label{eqforL1prime2}
\frac{\del \L_1'}{\del \tau}+\frac{\L_1'}{\tau_R} = -\frac{1}{\tau}\left[s_1 \L_0 + \left(a_1+t_1+\frac{c_1 s_2}{s_1}\right)\L_1' \right].
\end{equation}
That is, Eq.~(\ref{eqforL1prime2}) can be read as an equation for $\Pi-\phi$, decoupled from the $\M$-moments. It is tempting to look at the contribution of $\L_2'$ to the second order coefficients as a renormalization of the coefficient $a_1$
\begin{equation}\label{a1bar2}
a_1 \mapsto a_1+t_1+c_1 \frac{s_2}{s_1}, 
\end{equation}
somewhat analogous to (\ref{eq:a1prime}). There is, however, a profound difference with what is achieved in Eq.~(\ref{eq:a1prime}), where $a_1$ is adjusted so that the collisionless fixed point is accurately reproduced. Here, in Eq.~(\ref{a1bar2}), the correction can be viewed as a mass correction to the second-order transport coefficient $a_1$ (this correction indeed vanishes in the massless limit). In the case of massless particles, the correction (\ref{eq:a1prime}) to $a_1$ is sufficient to obtain an excellent agreement between kinetic theory and second-order hydrodynamics. However, in the massive case, as we have abundantly discussed in the previous section, a nonuniform renormalization is needed to account for the different behaviours of $\L_2$ in the vicinity of the collisionless and hydrodynamic fixed points. This is what is implemented in the three-moment truncation, and is shown in the previous section to yield excellent results. 

Having achieved the decoupling, we turn now to the equation for $\M_0$, which we write for $\M_0'=\M_0-\M_0^{\rm eq}$,
\begin{equation}\label{m0_treatment}
\pder{\M_0'}{\tau} + \frac{\M_0'}{\tR} = -\frac{1}{\tau} \left[ u_0 \L_0 + a_0 \M_0' +v_0 \L_1' +c_0 \M_1' \right].
\end{equation}
At this point, we could proceed as we did earlier for $\L_1'$, and exemplify another ambiguity present for the Bjorken flow. However, this is not necessary since our main goal has been achieved already with the study of $\L_1'$. We shall then use the  contributions of $\M_1'$ to $\Pi$ ($\Pi_{ \in \M_1'}$) and $\phi$ ($\phi_{ \in \M_1'}$) in Eq.~\eqref{PiinM1} in the previous section. By doing so, one keeps the coefficients $\delta_{\Pi\Pi}$ and $\lambda_{\Pi\phi}$ the same as in Eq.~\eqref{bulk_coeff}. Therefore, the evolution of bulk pressure is obtained as
\begin{equation}\label{newPieq}
\frac{d\Pi}{d\tau} + \frac{\Pi}{\tR} = -\frac{1}{\tau}\left[-\frac{u_0}{3}\L_0 + \delta_{\Pi\Pi}\Pi
-\lambda_{\Pi\phi} \phi\right],
\end{equation}
which is the same as \eqref{bulk_bj} since $\beta_\Pi= -\frac{u_0}{3}\L_0$.

Evolution of shear stress obtained from Eqs.~(\ref{eqforL1prime2}, \ref{newPieq}) is
\begin{align}
\frac{d\phi}{d\tau} +\frac{\phi}{\tR} =  \frac{1}{\tau} \bigg[& \!\left(s_1 c_0-\frac{u_0}{3} \right)\!\L_0 - \!\left(\! a_1+t_1+\frac{c_1s_2}{s_1} -\lambda_{\Pi\phi} \!\right)\! \phi 
\nn
&+ \left( a_1+t_1+\frac{c_1s_2}{s_1} -\delta_{\Pi\Pi} \right) \Pi  \bigg] .
\end{align}
Note that $\left(s_1c_0-u_0/3 \right)\L_0 = 4\beta_\phi/3$. At this point, we define the new transport coefficients (cp. Eq.~(\ref{shear_coeff}))
\begin{align}\label{newcoeffs}
a_1+t_1+\frac{c_1s_2}{s_1} -\lambda_{\Pi\phi}  &\equiv  \overline{\delta_{\phi\phi}+\tau_{\phi\phi}/3} ,
\\
\frac{3}{2}\left( a_1+t_1+\frac{c_1s_2}{s_1} -\delta_{\Pi\Pi} \right) &\equiv \overline{\lambda_{\phi\Pi}} ,
\end{align}
in terms of which Eq.~(\ref{shear_bj}) for $\phi$ takes the form
\begin{equation}\label{newphieq}
\frac{d\phi}{d\tau} +\frac{\phi}{\tR} = \frac{1}{\tau} \left[\frac{4}{3} \beta_\phi - \left(\overline{\delta_{\phi\phi}+\frac{\tau_{\phi\phi}}{3}} \right) \phi + \frac{2}{3} \overline{\lambda_{\phi\Pi}} \, \Pi  \right] .
\end{equation}

\begin{figure}[t!]
    \centering
    \includegraphics[width=\linewidth]{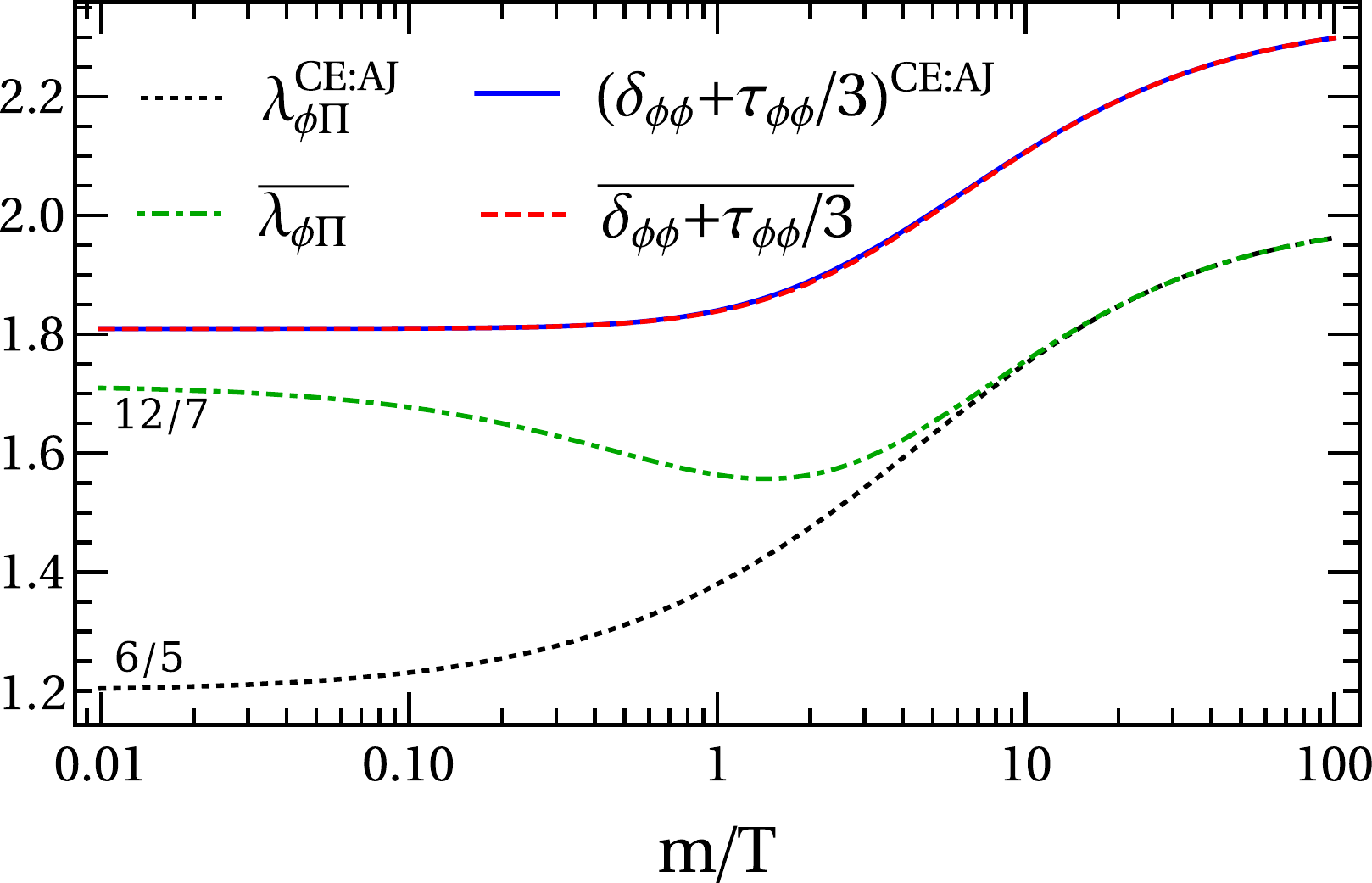}
    \vspace*{-.6cm}
    \caption{Comparison of second-order transport coefficients obtained in Ref.~\cite{Jaiswal:2014isa} with those obtained in Sec.~\ref{sec:newcoeff}. The coefficient $\tau_{\phi\phi}+\delta_{\phi\phi}/3$ obtained in Ref.~\cite{Jaiswal:2014isa} (solid blue) and in this work (red dashed) differ only slightly. However, the coefficient $\overline{\lambda_{\phi\Pi}}$  (dot-dashed green)  differs substantially in the small-mass limit from the value of $\lambda_{\phi\Pi}$ obtained in \cite{Jaiswal:2014isa} (black dotted line).}
    \label{Fig_newcoeff}
  \vspace*{-.2cm}
\end{figure}

The comparison of these new coefficients with the ones obtained previously is shown in Fig.~\ref{Fig_newcoeff}. As can be seen in this figure, the combination of coefficients $\delta_{\phi\phi}+\tau_{\phi\phi}/3$ is hardly modified, but the change in $\lambda_{\phi\Pi}$ can be substantial for  small values of $m/T$. The latter coefficient is important in understanding the effect of the bulk pressure in the evolution of the shear pressure; see Ref.~\cite{Denicol:2014mca} for a detailed study. We remark here that the change in the value of this coefficient has little effect on the hydrodynamical evolution for the initial conditions that we have considered. This is because the bulk pressure remains small during the evolution, and therefore the term $\lambda_{\phi\Pi} \Pi$ is negligible. 

Finally, it should be emphasized that the coefficients of the gradient series of $\Pi$ and $\phi$, that is the coefficients of the expansions of $\Pi$ and $\phi$ in inverse powers of $w=\tau/\tau_R$, remain unchanged in the reshuffling of the various second-order hydrodynamic quantities considered in this section (see Appendix~\ref{app:gr2}). A detailed analysis of the ambiguities that arise in second-order hydrodynamics,  and their implications, will be discussed in a separate work \cite{SunilJPB}.

\vspace*{-.2cm}
\subsection{Some results and discussion}
\vspace*{-.2cm}

\begin{figure}[t!]
    \centering
    \includegraphics[width=\linewidth]{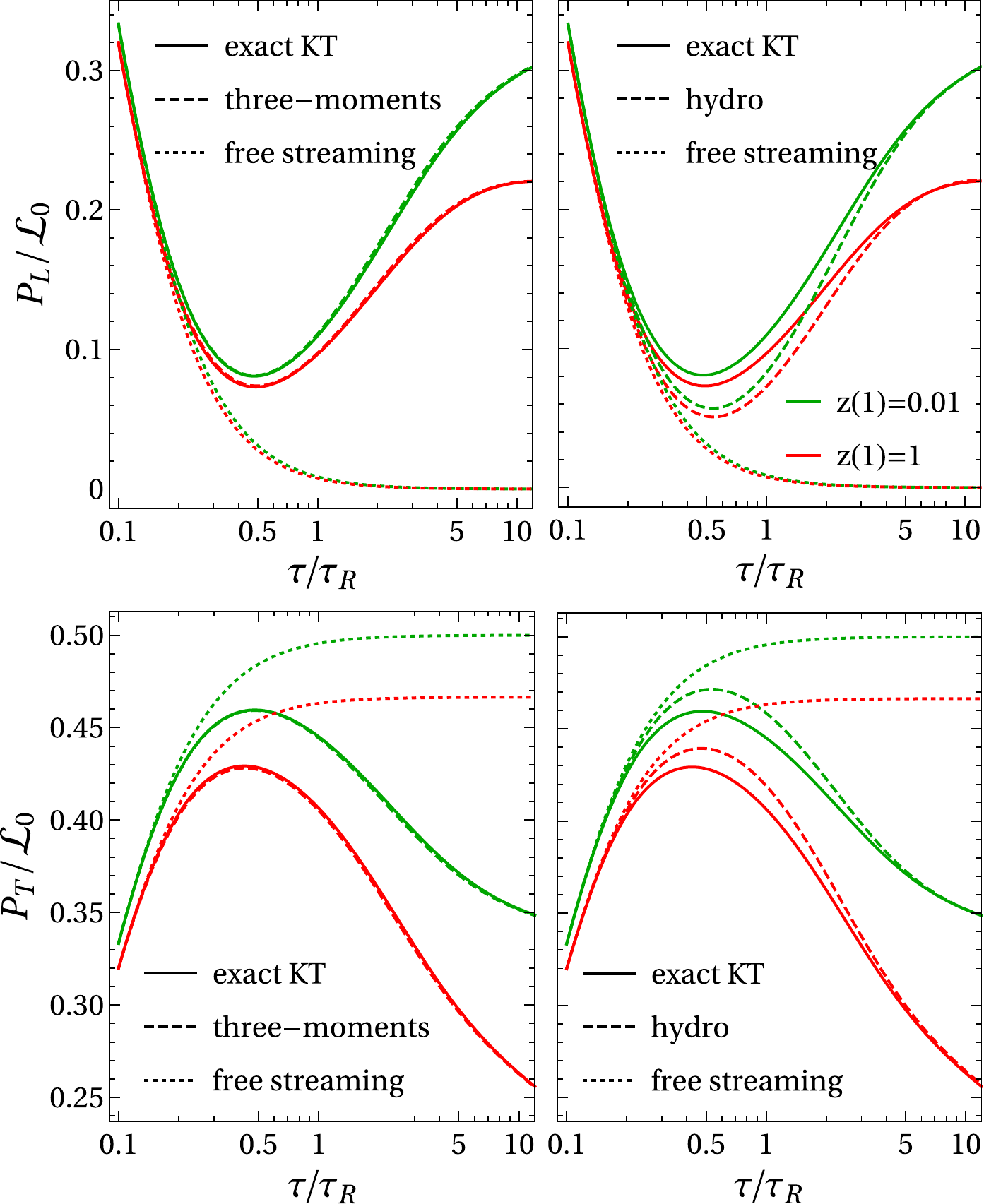}
    \vspace*{-.3cm}
    \caption{The longitudinal ($P_L$) and the transverse ($P_T$) pressures, normalized to the energy density $\L_0$,  as a function of time. The initial condition is set at $\tau_{\rm in} = 0.1\tau_R$, with an isotropic distribution function. Left panels:  three-moment truncation with interpolation~(\ref{ansatzM1_iso}, \ref{ansatzL2_iso})) (dashed lines), and exact kinetic theory (solid lines). Right panels: solutions of hydrodynamic Eqs.~(\ref{hydro_evol}) (dashed lines) and exact kinetic theory (solid lines). The dotted lines in all panels represent the (exact) free-streaming  evolution, reaching the free-streaming fixed point values at late times (for massless particles, $P_L\to 0$, $P_T\to 0.5 \L_0$). The red and green lines  correspond to $z(1)=1$ and $0.01$, respectively.}
    \vspace*{-.2cm}
    \label{fig:IC50}
\end{figure}

We end this section on hydrodynamics by comparing results obtained by solving the second-order hydrodynamic equations~\eqref{hydro_evol} with those obtained from the exact solution of the kinetic equation. Such a comparison is illustrated in Fig.~\ref{fig:IC50} for two physical quantities, the longitudinal and the transverse pressures, respectively $P_L$ and $P_T$. Also shown in Fig.~\ref{fig:IC50} is the approximate solution of the kinetic equation with the three-moments equations~\eqref{L_n:trunc} using the interpolation~(\ref{ansatzM1_iso}, \ref{ansatzL2_iso}). As can be seen in Fig.~\ref{fig:IC50}, left panels, the approximate solution is nearly indistinguishable from the exact solution. This is so in spite of the fact that, for the rather large initial time $\tau_{\rm in}=0.1\,\tau_R$ that is used here, the collisions play a role very early on and prevent the system to reach  the free-streaming fixed point  before  entering into the hydrodynamic regime (this is to be contrasted with the study presented in the previous section, see in particular Fig.~\ref{fig:IC40}, where the free-streaming fixed point is reached well before the system enters the hydrodynamic regime). Remarkably, as can be seen in the right panels of Fig.~\ref{fig:IC50}, the exact results are also well reproduced, albeit with less accuracy, by second-order hydrodynamics. In particular, a short free-streaming regime is seen in both the kinetic description and in the hydrodynamic one. There is of course nothing typically ``hydrodynamic'' here. As we have already emphasized, hydrodynamics  becomes a valid description only for times $\tau\gtrsim \tau_R$ (see, e.g., Fig.~\ref{fig:g0_exact}). As discussed in \cite{Blaizot:2021cdv} the origin of this particular behavior of ``second-order hydrodynamics'' may be traced back to the specificity of the Israel-Stewart approach, the time derivative of the viscous pressure capturing approximately some of the features of the collisionless regime of the expanding system. In fact, with the chosen initial conditions, we do not see the unphysical feature of the (uncorrected) Israel-Stewart approach: had we started the evolution with smaller values of the longitudinal pressure, one would have observed an excursion of the system into a region of negative longitudinal pressure (corresponding effectively to the free-streaming fixed point as approximated by the Israel-Stewart approach \cite{Blaizot:2019scw,Blaizot:2021cdv})%
    \footnote{Large deviations from kinetic theory may be observed in hydrodynamics at early time for initial conditions that yield large bulk pressures \cite{Jaiswal:2021uvv}.}.  
Of course this unphysical feature is cured in the three-moment truncation by using the simple interpolations presented in the previous section which, in the case of massless particles, amounts to a simple renormalization of a second-order transport coefficient \cite{Blaizot:2021cdv}. We have not considered this correction in the present case because that would mean essentially duplicating what we have done for the three-moment truncation which, as we have shown, captures accurately the dynamics from the collisionless to the  hydrodynamic regimes and involves essentially the same degrees of freedom as hydrodynamics.

\vspace*{-.4cm}
\section{Conclusions}
\label{sec_conclusions}
\vspace*{-.2cm}

In this paper we have extended the analysis of Refs.~\cite{Blaizot:2017lht,Blaizot:2017ucy,Blaizot:2019scw,Blaizot:2021cdv}
to the case of massive particles. We have shown that the two moments of the distribution function, $\L_0$ and $\L_1$, that represent in the massless case, the energy density and the shear pressure, must be complemented in the massive case by a third moment, $\M_0$,  equal to the trace of the energy-momentum tensor. We have shown that the equations for $\L_0$ and $\L_1$ decouple from that of $\M_0$ and can be solved independently in order to get, in particular, the energy density, needed as input to solve the equation for $\M_0$. The fixed-point structure that determines the dynamics in the massive case emerges entirely from the equations for $\L_0$ and $\L_1$ which can be transformed into a nonlinear differential equation similar to that obtained in the massless case. The free-streaming fixed points are unchanged, while the hydrodynamic fixed point is modified to take into account the dependence of the speed of sound on the equation of state. The vicinity of the hydrodynamic fixed point is described by the Navier-Stokes equation with viscosities that can be deduced in leading order from the Chapman-Enskog expansion. In this work, we have assumed that the mass of the particles is a constant and have restricted ourselves to situations where the evolution is stopped at times where the ratio $m/T$ remains sufficiently small to avoid entering too deeply into the non-relativistic regime.

As was the case for massless particles, a simple renormalization, whose goal is to correct for the neglect of the infinite tower of higher moments, allows us to  reproduce accurately the exact result of kinetic theory within a  three-moment truncation. This three-moment truncation contains  as a limiting case the hydrodynamic equations. We have discussed in particular how the second-order hydrodynamic equations emerge from the three-moment truncation, and pointed out potential ambiguities in the definition of some of the second-order transport coefficients. These ambiguities are connected, in part, with a possible decoupling of the equations of second-order hydrodynamics that, to the best of our knowledge, was never considered before.

\acknowledgments

The authors thank Chandrodoy Chattopadhyay for helpful discussions and insightful comments. This work was started when RSB visited IPhT, Saclay in the summer of 2019. RSB and SJ acknowledge the kind hospitality of IPhT, Saclay and NISER, Bhubaneswar where parts of this work were completed. RSB and AJ would like to thank the Department of Science and Technology, India for the research grant no. SERB/CRG/2019/000807. RSB and AJ acknowledge the support of the IFCPAR/CEFIPRA-funded project no. 6403. ZC and LY are partly supported by National Natural Science Foundation of China through grant No. 11975079.


\appendix

\vspace*{-.2cm}
\section{Exact solution of the RTA Boltzmann equation } 
\label{app:exactKT_sol}
\vspace*{-.2cm}

The solution of the kinetic equation~(\ref{keq_2}) can be formally written as \cite{Baym:1984np, Florkowski:2013lya, Florkowski:2014sfa}:
\begin{align}\label{soln}
f(\tau; p_\perp, p_z) =& D(\tau,\tau_{\rm in}) f_\mathrm{in}(p_\perp, p_z \tau/\tau_{\rm in})
\nn
&+ \int_{\tau_{\rm in}}^{\tau} \, \frac{d\tau'}{\tau_R(\tau')} \, D(\tau,\tau') \, f_\mathrm{eq} (p_\perp, p_z \tau/\tau'),
\end{align}
where 
\begin{equation}
D(\tau_2,\tau_1) = {\rm e}^{-(\tau_2-\tau_1)/\tau_R}.
\end{equation}
Here $f_\mathrm{in}$ is the initial distribution function at proper time $\tau_{\rm in}$ and $f_\mathrm{eq} $ is the equilibrium distribution, taken to be a Boltzmann distribution at a local temperature $T(\tau)$ that is determined by the Landau matching condition:
\begin{align}\label{LM:1}
\int_\p\! \sqrt{m^2+p^2} \, f(\tau; p_\perp, p_z) 
=\! \int_p\! \sqrt{m^2+p^2}\, \exp\!\left(\!- \frac{\sqrt{m^2+ p^2}}{T(\tau)}\! \right).
\end{align}
Together with Eq.~(\ref{soln}) for $f$,  Eq.~(\ref{LM:1}) allows the numerical determination of the temperature at each time, via an iterative procedure \cite{Banerjee:1989by, Florkowski:2013lya, Florkowski:2014sfa}. To proceed, it is convenient to express all momenta in units of the mass $m$, and solve  Eq.~\eqref{LM:1} for the ratio $z(\tau) \equiv m/T(\tau)$:
\begin{align}\label{LM:2}
\int_{\bar{\p}}\! \sqrt{1+\bar{p}^2}\, f(\tau; \bar{p}_\perp, \bar{p}_z) 
= \int_{\bar{\p}}\! \sqrt{1+\bar{p}^2}\, \exp\left(- z(\tau ) \sqrt{1+ \bar{p}^2} \right),
\end{align}
where $p_i =m\, \bar{p_i}$ and $\int_{\bar{\p}} \equiv \int \frac{{\rm d}^3\bar{p}}{(2\pi)^3}$. Note that we do not need to specify the initial mass and  temperature, but only the ratio $z(\tau_{\rm in})$. 

Once $z(\tau)$ has been obtained, we can use it to calculate the normalized moments $\L_n$ and $\M_n$. For instance, $\L_n$ is given by 
\begin{align}\label{exact_ln}
&\frac{\L_n(\tau)}{m^4} = D(\tau,\tau_0) \!\int_{\bar{\p}}\! \sqrt{1+\bar{p}^2} \, P_{2n}\!\left(\!\frac{\bar{p}_z}{\bar{p}_0}\!\right) f_\mathrm{in}\!\left(\!\bar{p}_\perp, \frac{\bar{p}_z \tau}{\tau_{\rm in}}\!\right) 
\nn
&+ \int_{\tau_{\rm in}}^{\tau}\! \frac{d\tau'}{\tau_R(\tau')} D(\tau,\tau') \!\int_{\bar{\p}}\! \sqrt{1+\bar{p}^2} \, P_{2n}\!\left(\!\frac{\bar{p}_z}{\bar{p}_0}\!\right) f_\mathrm{eq}\!\left(\!\bar{p}_\perp, \frac{\bar{p}_z \tau}{\tau'}\!\right),
\end{align}
and a similar equation holds for $\M_n$.

In order to provide  more details on the numerics, consider the attractor solutions discussed in Section~\ref{sec_trunc_FP}. The initial distribution is chosen to describe the vicinity of the stable collisionless fixed point, i.e., it is peaked in the $z$ direction. We use the simple function 
\begin{equation}\label{f_in}
f_\mathrm{in}(p_\perp, p_z) = \exp\left(- \frac{\sqrt{m^2+ p_\perp^2 + \xi^2 p_z^2 }}{\Lambda} \right).
\end{equation}
where the parameter $\xi$ mimics a free-streaming evolution till the initial time, and $\Lambda$ characterizes the typical momentum scale (see the next Appendix).  We choose $\xi^2=10^4$, and $m/\Lambda=\{ 1.38\times 10^{-4}, 6.88\times 10^{-3}, 1.41\times 10^{-2}, 1.115\times 10^{-1}\}$ at $\tau_{\rm in} = 1\times 10^{-5}$\, fm/$c$ to obtain $z(\tau)$ from the Landau matching condition~\eqref{LM:2} for the green, blue, red, and orange curves, respectively ($m/\Lambda$ is tuned to get the desired value of $m/T$ at $\tau=\tau_R$). We set $w=\tau/\tau_R$.   Starting from an initial guess for $z(w)$ on $300$ nonuniform grid points in $w$ in the interval $\{ w_{\rm in}, 600\}$, we numerically iterate till the maximum error (evaluated at these $300$ grid points) between two subsequent iterations is $\leq0.01\%$ for $z(w)$, i.e., max$\big|\frac{z_n-z_{n-1}}{z_n}\big|$ ($n$ represents the $n^{\rm th}$ iteration) evaluated at these $300$ grid points is $\leq 10^{-4}$. We then evaluate $\L_n$  using Eq.~(\ref{exact_ln}) and similarly for $\M_n$.

\section{More on initial conditions} 
\label{app:B}

In this appendix, we comment on the role of the mass and of the anisotropy  of the initial momentum distribution of the particles. 

Let us first consider an isotropic distribution. We assume that such a distribution is characterized by a single momentum scale $\Lambda$ (e.g., $\Lambda$ can be the saturation momentum $Q_s$; alternatively, one may associate $\Lambda$ to the initial effective temperature, making use of the Landau matching condition). In other words, we assume that $f_{\rm in}=f_{\rm in}(p_0/\Lambda)$ where $p_0=\sqrt{p^2+m^2}$ is the energy of a particle. 

With such a distribution, the energy density and the trace of the energy-momentum tensor are easily evaluated. We have
\begin{equation}
\varepsilon_{\rm in}(m)=\int_\p \sqrt{m^2+p^2} f_{\rm in}(p_0/\Lambda)\equiv\Lambda^4 \,g(m/\Lambda),
\end{equation}
and 
\begin{align}
\M_0^{\rm in}(m) &= m^2 \int_\p \frac{1}{\sqrt{m^2+p^2}} f_{\rm in}(p_0/\Lambda)
\equiv m^2\Lambda^2\, h(m/\Lambda),
\end{align}
where $g$ and $h$ are dimensionless functions of $m/\Lambda$, which have finite limits when $m\to 0$. All the non-trivial mass dependence is contained in these functions. 

These expressions of $\varepsilon_{\rm in}$ and $\M_0^{\rm in}$ show that when $m\ll\Lambda$, $\M_0^{\rm in}$ is parametrically much smaller than the energy density. In the opposite limit, $m\gg \Lambda$, the system is in a non-relativistic regime and 
\begin{align}
\varepsilon_{\rm in}(m\gg \Lambda)\approx& \int_\p \left( m+\frac{p^2}{2m}\right) f_{\rm in}(p_0/\Lambda)
= n_{\rm in} m +\frac{3}{2} P_{\rm in},
\end{align}
and 
\begin{equation}
\M_0^{\rm in}(m\gg \Lambda)\approx m \int_\p f_{\rm in}(p_0/\Lambda)=n_{\rm in} m.
\end{equation}
We have set
\begin{equation}
n_{\rm in}=\int_\p f_{\rm in}(p_0/\Lambda),
\end{equation}
and $P_{\rm in}$ is the initial pressure. In this limit, $\M_0^{\rm in}\lesssim \varepsilon_{\rm in}$.  The transition to this non-relativistic regime occurs when $m\sim \Lambda$. As already mentioned in Section~\ref{sec:kinetic}, we are not interested in this unphysical regime, so we shall assume that $m\ll \Lambda$. 

Some of the numerical calculations presented in this paper involve an anisotropic initial distribution of the form (\ref{f_in}) where $\xi$ controls the anisotropy (a large value of $\xi$ corresponding to a distribution that is peaked in the $z$ direction). We assume $m\ll\Lambda$. Without too much effort, one can estimate the parametric dependence of various physical quantities on the parameters $\Lambda$ and $\xi$. In the limit of a large anisotropy (required to initialize the flow near the stable free-streaming fixed point), $\xi$ is large. It provides then a cutoff on the $p_z$ integration of order $\Lambda/{\xi}$. The transverse momentum is cut off at the scale $\Lambda$. One can then easily estimate the initial energy density and the moment $\M_0^{\rm in}$. We get
\begin{equation}
\varepsilon_{\rm in}\simeq \frac{1}{12}\frac{\Lambda^4}{ \xi},\qquad \M_0^{\rm in}\simeq \frac{m^2}{4\pi^2}\frac{\Lambda^2}{\xi}.
\end{equation}
The energy density can also be estimated from an equilibrium distribution with temperature $T_{\rm in}$, and Landau matching then allows us to determine $T_{\rm in}$. We get 
\begin{equation}
\varepsilon_{\rm in}\simeq \frac{T_{\rm in}^4}{8\pi^2}, \qquad 
\frac{T_{\rm in}}{\Lambda}\simeq \left( \frac{2\pi^2}{3 \xi} \right)^{1/4}\simeq   \frac{1}{\xi^{1/4}}.
\end{equation}
Similarly, we can calculate the corresponding equilibrium value of $\M_0$. We get
\begin{equation}
M_0^{\rm eq}\simeq\frac{m^2 T_{\rm in}^2}{8\pi^2},\qquad \frac{\M_0^{\rm in}}{\M_0^{\rm eq}}\simeq\frac{1}{\xi^{1/2}}.
\end{equation}
These simple estimates are useful to understand the order of magnitude of quantities discussed in Section~(\ref{sec_FP_mn}).

\vspace*{-.2cm}
\section{Chapman-Enskog expansion}
\label{app:CE_exp}
\vspace*{-.2cm}

The deviation of the distribution function $f$ from its equilibrium value $f^{\rm eq}(p_0/T)$ at late time can be calculated using the  Chapman-Enskog expansion. In leading order, this is given by (see, e.g., Ref. \cite{Blaizot:2019scw})
\begin{equation}\label{app_CE_ideal}
\delta f^{(1)} = \frac{\tau_R}{\tau} \left(\frac{p_z^2}{p_0^2} -c_s^2 \right)  p_0\frac{\del f^{\rm eq}}{\del p_0} \,,
\end{equation}
where we have used the leading-order relation 
\begin{equation}
\frac{d\ln T}{d\ln \tau} = - c_s^2 \,.
\end{equation}

Using this expression of $\delta f^{(1)} $, one can easily calculate the leading-order contributions to the bulk and shear pressures. One finds
\begin{align}
\Pi &=\frac{1}{3} \int_\p \frac{p^2}{p_0}\delta f^{(1)}
\nn
&= \frac{\tau_R}{3\tau}\left\{\left( \frac{1}{3}-c_s^2 \right) \int_\p\frac{\del f^{\rm eq}}{\del p_0} \frac{p^4}{p_0^2}-m^2 c_s^2 \int_\p\frac{\del f^{\rm eq}}{\del p_0} \frac{p^2}{p_0^2}  \right\},
\end{align}
and 
\begin{equation}
\phi=-\frac{2}{3}\int_\p \left(\frac{p_z^2}{p_0}-\frac{p_\perp^2}{2p_0}  \right)\delta f^{(1)}=-\frac{4}{45}\frac{\tau_R}{\tau}\int_\p\frac{\del f^{\rm eq}}{\del p_0} \frac{p^4}{p_0^2}.
\end{equation}
We have used the shorthand notation for the three-momentum integration  $\int_\p\equiv \int\frac{\rmd^3p}{(2\pi)^3}$. The momentum integrals can be reduced to integrals over the equilibrium distribution function by integrating by parts, after making the change of variables $p=m\sinh\theta$ and $p_0=m\cosh\theta$.  We get
\begin{align}
\int_\p\frac{\del f^{\rm eq}}{\del p_0} \frac{p^4}{p_0^2}& = \int_\p f^{\rm eq}\left( \frac{p^4}{p_0^3}-5\frac{p^2}{p_0} \right),
\nn
\int_\p\frac{\del f^{\rm eq}}{\del p_0} \frac{p^2}{p_0^2} &= -\int_\p f^{\rm eq}\left( \frac{2}{p_0}+\frac{m^2}{p_0^3} \right).
\end{align}
These integrals can in turn be expressed in terms of the equilibrium values of the moments. We have indeed
\begin{align}
\M_0^{\rm eq} &= \int_\p  \frac{m^2}{p_0}f^{\rm eq}=\varepsilon-3P,
\nn
\M_1^{\rm eq} &= -\frac{1}{2} \int_\p \frac{m^4}{p_0^3}f^{\rm eq}, 
\nn
\L_2^{\rm eq} &= -\frac{1}{2}\M_0^{\rm eq}-\frac{7}{4}\M_1^{\rm eq},
\end{align}
so that
\begin{equation}
\int_\p\frac{\del f^{\rm eq}}{\del p_0} \frac{p^4}{p_0^2}=-3(\varepsilon+P)+2(\M_0^{\rm eq}-\M_1^{\rm eq}),
\end{equation}
and
\begin{equation}
\int_\p\frac{\del f^{\rm eq}}{\del p_0} \frac{p^2}{p_0^2}=-\frac{2}{m^2} (\M_0^{\rm eq}-\M_1^{\rm eq}).
\end{equation}
Combining all these results, one obtains the leading-order contributions to $\Pi$ and $\phi$ in the form
\begin{align}
\Pi &= \frac{\tau_R}{3\tau}\left[(3c_s^2-1)(\varepsilon+P)+\frac{2}{3} \left(\M_0^{\rm eq}-\M_1^{\rm eq} \right) \right],
\nn
\phi &= \frac{\tau_R}{\tau}\left[ \frac{16}{45}\varepsilon-\frac{4}{15}\M_0^{\rm eq}+\frac{8}{45} \M_1^{\rm eq}  \right].
\end{align}
One verifies easily that these expressions agree with those given in the main text (see Eqs.~\eqref{eq:PiLO} and \eqref{eq:PhiLO}). In order to verify directly the expression of $\L_1'$ given in the main text, it is useful to recall the Landau matching condition $\int_\p p_0 \delta f^{(1)}=0$, which is easily verified:
\begin{align}
\int_\p p_0 \delta f^{(1)} &= \frac{\tau_R}{\tau}\int_\p \frac{\del f^{\rm eq}}{\del p_0}  p_0^2\left( \frac{p^2}{3p_0^2}-c_s^2  \right)
\nn
&= -\frac{\tau_R T}{\tau}\left[ \frac{\del P}{\del T} - c_s^2 \frac{\del \varepsilon}{\del T} \right]=0.
\end{align}
We have used $\rmd P/ \rmd\varepsilon =c_s^2$ and 
\begin{equation}\label{eq:dfdp0}
 p_0\frac{\del f^{\rm eq}}{\del p_0}=-T\frac{\del f^{\rm eq}}{\del T}.
\end{equation}
To proceed systematically in forthcoming calculations, it proves convenient to set
\begin{equation}
 p^2=p_z^2+p_\perp^2,\quad q^2=p_z^2-\frac{p_\perp^2}{2}, \quad p_z^2=\frac{1}{3}(p^2+2 q^2).
\end{equation}
Then, 
\begin{equation}\label{eq:deltaf1}
\delta f^{(1)}=\frac{\tau_R}{\tau}\frac{1}{p_0}\frac{\del f^{\rm eq}}{\del p_0}\left[ \left(\frac{p^2}{3}-c_s^2 p_0^2\right)+\frac{2}{3}q^2\right]=\mathcal{B}+\mathcal{S}q^2.
\end{equation}
The calculation of $\L_1'$ goes then as follows ($P_2(x)=(3x^2-1)/2)$
\begin{align}
\L_1'&=\frac{1}{2}\int_\p p_0\left(3\frac{p_z^2}{p_0^2}-1  \right) \delta f^{(1)}\nonumber\\
&=\frac{1}{2}\int_\p \frac{1}{p_0} (p^2+2q^2)(\mathcal{B}+\mathcal{S}q^2)
= \frac{1}{2}\int_\p \frac{1}{p_0} (\mathcal{B}p^2+2 \mathcal{S} q^4)
\nn
&=\frac{\tau_R}{2\tau}\int_\p \frac{\del f^{\rm eq}}{\del p_0} \frac{1}{p_0^2}\left[ \left(\frac{p^2}{3}-c_s^2 p_0^2\right)p^2+\frac{2}{3}q^4  \right]
\nn
&=\frac{\tau_R}{2\tau}\int_\p \frac{\del f^{\rm eq}}{\del p_0} \left[ \frac{p^4}{p_0^2}\left(\frac{1}{3}-c_s^2\right)-m^2 c_s^2 \frac{p^2}{p_0^2}+\frac{2}{15}\frac{p^4}{p_0^2}  \right],
\end{align}
where in going to the second line, we have used the Landau matching condition, and in the last line we have used $\int_\p q^4 f(p)=\frac{1}{5}\int_\p p^4 f(p)$. One recognizes the integrals that contribute to $\Pi$ and $\phi$, respectively, and then, remembering that $\L_1'=(\Pi-\phi)/c_0$ one recovers the expression of $\L_1'$ given in the main text (Eq.~(\ref{L1prime})).

In order to determine the second-order transport coefficients, we need the expressions of $\L_2'$ and $M_1'$. Consider first $\M_1'$. We have
\begin{align}\label{eq:M1p0}
\M_1'&=\frac{1}{2}\int_\p \frac{m^2}{p_0}\left(3\frac{p_z^2}{p_0^2}-1  \right) \delta f^{(1)}
\nn
&=\frac{m^2}{2}\int_\p \frac{1}{p_0^3}\left(p^2+2q^2-p_0^2 \right) \delta f^{(1)}\nonumber\\
&=\frac{m^2}{2}\int_\p \frac{1}{p_0^3}\left[ -m^2 \mathcal{B}+2\mathcal{S}q^4 \right]
\nn
&=\frac{m^2}{2}\frac{\tau_R}{\tau}\!\int_\p\! \frac{\del f^{\rm eq}}{\del p_0}\! \left[ -\frac{m^2p^2 }{p_0^4} \!\left(\frac{1}{3}-c_s^2\right)\! + c_s^2 \frac{m^4 }{p_0^4}+\frac{4}{15}\frac{p^4}{p_0^4} \right].
\end{align}
There are new integrals to evaluate. The first one is  
\begin{equation}\label{eq:p2p04}
\int_\p \frac{\del f^{\rm eq}}{\del p_0}\frac{p^2}{p_0^4}=-3m^2\int_\p \frac{f^{\rm eq}}{p_0^5} .
\end{equation}
The  integral in the last term in Eq.~(\ref{eq:M1p0}) can be reduced by using $p^2=p_0^2-m^2$. We get
\begin{align}
\int_\p \frac{\del f^{\rm eq}}{\del p_0}\frac{p^4}{p_0^4} 
&= -\int_\p f^{\rm eq} \left(5\frac{p^2}{p_0^3}-3\frac{p^4}{p_0^5} \right) 
\nn
&=-\int_\p f^{\rm eq}\left( \frac{2}{p_0}+\frac{m^2}{p_0^3}-3\frac{m^4}{p_0^5} \right),
\end{align}
so that
\begin{equation}
m^2\int_\p \frac{\del f^{\rm eq}}{\del p_0}\frac{p^4}{p_0^4}=-2(\M_0^{\rm eq}-\M_1^{\rm eq})+3m^6 \int_\p\frac{f^{\rm eq}}{p_0^5}.
\end{equation}
The integral in the middle term in the last line of Eq.~(\ref{eq:M1p0}), 
\begin{align}
\int_\p \frac{\del f^{\rm eq}}{\del p_0}\frac{m^4}{p_0^4}
&=m^2\int_\p \frac{\del f^{\rm eq}}{\del p_0}\frac{p_0^2-p^2}{p_0^4}
\nn
&=\int_\p \frac{\del f^{\rm eq}}{\del p_0}\left(\frac{m^2}{p_0^2}-\frac{m^2p^2}{p_0^4}\right).
\end{align}
The last of the two integrals in this expression is known (see Eq.~(\ref{eq:p2p04})). To calculate the first integral we use (\ref{eq:dfdp0}) and get 
\begin{equation}
\int_\p \frac{\del f^{\rm eq}}{\del p_0} \frac{m^2}{p_0^2}=- T\frac{\del }{\del T}\int_\p f^{\rm eq}\frac{m^2}{p_0^3}=\frac{2T}{m^2}\frac{\del \M_1^{\rm eq}}{\del T}.
\end{equation}
By combining these previous results we get
\begin{equation}\label{m1pCE}
\M_1' =\frac{\tau_R}{\tau} \left[ \frac{9m^6}{10}\int_\p \frac{f^{\rm eq}}{p_0^5} +c_s^2 T\frac{\del \M_1^{\rm eq}}{\del T} -\frac{4}{15} (\M_0^{\rm eq}- \M_1^{\rm eq}) \right].
\end{equation}

Consider next $\L_2'$. We have (with $P_4(x)=(3-30 x^2+35 x^4)/8$).
\begin{equation}
\L_2'=\int_\p \frac{p_0}{8} \left(3 -30\frac{p_z^2}{p_0^2} +35\frac{p_z^4}{p_0^4} \right) (\mathcal{B}+\mathcal{S}q^2).
\end{equation}
The first term ($\sim 3 p_0$) does not contribute because of the Landau matching condition. We are left with
\begin{align}
    \L_2'&=\frac{1}{8}\int_\p \left[ -\frac{10}{p_0}(p^2+2 q^2) +\frac{35}{9 p_0^3}(p^2+2q^2)^2 \right] (\mathcal{B}+\mathcal{S}q^2)
    \nn
    &=\frac{1}{8}\int_\p \bigg\{ \mathcal{B} \left[-10\frac{p^2}{p_0} +\frac{35}{9 p_0^3}(p^4+4q^4 ) \right]
    \nn
    &\qquad \quad+\mathcal{S}\left[ -20\frac{q^4}{p_0}+\frac{140}{9} \frac{p^2q^4}{p_0^3}+\frac{140}{9} \frac{q^6}{p_0^3} \right]\bigg\}.
\end{align}
We shall evaluate separately the two contributions from $\mathcal{B}$ and $\mathcal{S}$. We have
{\small
\begin{align}
    \L_{2\mathcal{B}}' &= \frac{1}{8}\int_\p \mathcal{B}\left[-10\frac{p^2} {p_0} +\frac{35}{9 p_0^3}(p^4+4q^4 )  \right]
    \nn
    &= \frac{\tau_R}{8\tau}\int_\p \frac{\del f^{\rm eq}}{\del p_0} \left(\frac{p^2}{3}-c_s^2 p_0^2  \right)\left[-10\frac{p^2}{p_0^2} +\frac{35}{9 p_0^4}\left(p^4+\frac{4}{5}p^4 \right) \right]
    \nn
    &= \frac{\tau_R}{8\tau}\int_\p \frac{\del f^{\rm eq}}{\del p_0} \left\{ \left[p^2\left(\frac{1}{3}-c_s^2   \right)-m^2 c_s^2\right]\left[-10\frac{p^2}{p_0^2} +7 \frac{p^4}{p_0^4} \right] \right\}
    \nn
    &= \frac{\tau_R}{8\tau}\int_\p \frac{\del f^{\rm eq}}{\del p_0} \bigg\{ \left(\frac{1}{3}-c_s^2   \right)\left[-10\frac{p^4}{p_0^2} +7 \frac{p^6}{p_0^4}  \right]
    \nn
    &\qquad\qquad\qquad\quad -m^2 c_s^2 \left[-10\frac{p^2}{p_0^2} +7 \frac{p^4}{p_0^4}  \right] \bigg\}
    \nn
    &= \frac{\tau_R}{8\tau}\int_\p \frac{\del f^{\rm eq}}{\del p_0} \bigg\{ \left(\frac{1}{3}-c_s^2 \right)\left[-3\frac{p^4}{p_0^2} -7 m^2\frac{p^4}{p_0^4}  \right]
    \nn
    &\qquad\qquad\qquad\quad -m^2 c_s^2 \left[-10\frac{p^2}{p_0^2} +7 \frac{p^4}{p_0^4}  \right] \bigg\}.
\end{align}}%
By using the expressions for the momentum integrals given above, we finally get
\begin{align}\label{jpbL2Ap}
    \L_{2\mathcal{B}}'&= \frac{\tau_R}{8\tau}\bigg[ 9(\varepsilon+P) \left(\frac{1}{3}-c_s^2\right) \!+\! \left(\frac{8}{3}-14 c_s^2\right) \left(\M_0^{\rm eq}-\M_1^{\rm eq}\right)
    \nn
    &\qquad\qquad -7m^6 \int_\p\frac{f^{\rm eq}}{p_0^5} \bigg]
    \nn
    &= \frac{\tau_R}{8\tau}\bigg[ 3(\varepsilon+P) +\frac{8}{3} (\M_0^{\rm eq}-\M_1^{\rm eq})-7m^6 \int_\p\frac{f^{\rm eq}}{p_0^5} 
    \nn
    &\qquad\qquad - c_s^2 \left\{ 9(\varepsilon+P) +14 (\M_0^{\rm eq}-\M_1^{\rm eq}) \right\}  \bigg]
    \nn
    &=\frac{\tau_R}{\tau} \left[ c_s^2 T \frac{\partial \L_2^{\rm eq}}{\partial T} + \frac{1}{3} (\M_0^{\rm eq}-\M_1^{\rm eq}) -\frac{7m^6}{8} \int_\p \frac{f^{\rm eq}}{p_0^5} \right].
\end{align}
For the $\mathcal{S}$ contribution, we get
\begin{align}
    \L_{2\mathcal{S}}' &= \frac{1}{8}\int_\p \mathcal{S}\left[ -20\frac{q^4}{p_0}+\frac{140}{9} \frac{p^2q^4}{p_0^3}+\frac{140}{9} \frac{q^6}{p_0^3} \right]
    \nn
    &= \frac{1}{8}\int_\p \mathcal{S}\left[ -4\frac{p^4}{p_0}+\frac{28}{9} \frac{p^6}{p_0^3}+\frac{8}{9} \frac{p^6}{p_0^3} \right]
    \nn
    &= \frac{1}{8}\int_\p \mathcal{S}\left[ -4\frac{p^4}{p_0}+4 \frac{p^6}{p_0^3} \right],
\end{align}
where we have used $\int_\p q^6 f(p)=\frac{2}{35}\int_\p p^6 f(p) $. We have therefore
\begin{align}
    \L_{2\mathcal{S}}' &= \frac{\tau_R}{12 \tau}\int_\p \frac{\del f^{\rm eq}}{\del p_0} \left[ -4\frac{p^4}{p_0^2}+4 \frac{p^6}{p_0^4} \right]=-\frac{m^2 \tau_R}{3\tau}\int_\p \frac{\del f^{\rm eq}}{\del p_0}\frac{p^4}{p_0^4}
    \nn
    &= \frac{\tau_R}{3 \tau} \left[ 2(\M_0^{\rm eq}-\M_1^{\rm eq})-3m^6 \int_\p \frac{f^{\rm eq}}{p_0^5}\right].
\end{align}

Finally,
\begin{equation}\label{L2prime}
    \L_2' =\frac{\tau_R}{\tau} \left[ c_s^2 T \frac{\partial \L_2^{\rm eq}}{\partial T} + \M_0^{\rm eq}-\M_1^{\rm eq} -\frac{15m^6}{8} \int_\p \frac{f^{\rm eq}}{p_0^5} \right] .
\end{equation}

The expressions given above for $\M_1'$ and $\L_2'$ are not unique, as is clear from the expressions given in Eq.~(\ref{jpbL2Ap}) for $\L_{2\mathcal{B}}'$. They can be transformed by using various identities. For instance, we have
\begin{equation}
    T \frac{\partial \L_2^{\rm eq}}{\partial T} = \frac{3}{8} T \frac{\partial\varepsilon}{\partial T} - \frac{9}{8} (\varepsilon+P) -\frac{7}{4} (\M_0^{\rm eq}-\M_1^{\rm eq}),
\end{equation}
which is easily derived by using $\L_2^{\rm eq}=-\frac{1}{2}\M_0^{\rm eq}-
\frac{7}{4}\M_1^{\rm eq}$ together with the following identities
\begin{equation}
T\frac{\del }{\del T} \left(\M_1^{\rm eq}+\frac{\M_0^{\rm eq}}{2}\right)=\M_0^{\rm eq}-\M_1^{\rm eq},
\end{equation}
and
\begin{equation}
T\frac{\del P}{\del T}=\varepsilon+P=Ts,\qquad s=\frac{\rmd P}{\rmd T}=c_s^2 \frac{\rmd \varepsilon}{\rmd T}.
\end{equation}
From these last two relations, we also get
\begin{equation}
T \frac{\partial \M_1^{\rm eq}}{\partial T} = -\frac{T}{2} \frac{\partial\varepsilon}{\partial T} + \frac{3}{2} (\varepsilon+P) + \M_0^{\rm eq}-\M_1^{\rm eq}.
\end{equation}

\vspace*{-.2cm}
\section{Gradient expansion} \label{app:gr}
\vspace*{-.2cm}

In this Appendix, we compare the coefficients of the gradient series for $\Pi$ and $\phi$ obtained on the one hand from the moment equations (\ref{L_n:trunc}), and on the other hand from the second-order hydrodynamic equations (\ref{hydro_evol}). We show that the values of these coefficients agree up to second order, independently of the choice made for the second-order transport coefficients discussed in Section~\ref{sec:newcoeff}. 

\vspace*{-.2cm}
\subsection{Moment equations} \label{app:gr1}
\vspace*{-.2cm}

In order to study the hydrodynamic regime, we rewrite the moment equations in terms of the  deviations $\L_n'$ of the moments $\L_n$ from their their equilibrium values $\L_n^{\rm eq}$, with  $\L_n'\equiv \L_n-\L_n^{\rm eq}$ for $n>0$ ($\L_0'= 0$).  
The $\L_n$-moment equations can then be recast in the form 
\begin{align}\label{Lnp_mom}
\pder{\L_0}{\tau} =& -\frac{1}{\tau} \left[\bar{a}_0 \L_0 + c_0 \L_1' \right], 
\nn
\pder{\L_n'}{\tau} =& -\frac{1}{\tau} \left[ s_n\, \L_0 + t_n\, \L_1' + a_n \L_n' + b_n \L_{n-1}' + c_n \L_{n+1}'\right] 
\nn
&- \frac{\L_n'}{\tR} , \quad \forall n \geq 1,
\end{align}
where $\bar{a}_0=\left( a_0 + c_0 \frac{\L_1^{\rm eq}}{\L_0} \right) = 1+P/\varepsilon$. The  coefficients $s_n,\, t_n$ are functions of $m/T$, and are given by
\begin{align}
s_n &= a_n \frac{\L_n^{\rm eq}}{\L_0} + b_n \frac{\L_{n-1}^{\rm eq}}{\L_0} + c_n \frac{\L_{n+1}^{\rm eq}}{\L_0} -\bar{a}_0 \pder{\L_n^{\rm eq}}{\varepsilon} ,
\nn
t_n &= -c_0  \pder{\L_n^{\rm eq}}{\varepsilon}.
\end{align}
To obtain these equations, we have used the following relation
\begin{equation}\label{eqLeqepsilon}
\pder{\L_n^{\rm eq}}{\tau}	= -\frac{1}{\tau} \left[ \left( a_0 + c_0 \frac{\L_1^{\rm eq}}{\L_0} \right) \pder{\L_n^{\rm eq}}{\varepsilon} \L_0 + \left(c_0 \pder{\L_n^{\rm eq}}{\varepsilon} \right) \L_1' \right],
\end{equation}
 which follows from the fact that the equilibrium values of the moments can be considered as functions of the energy density, from which they acquire their time dependence. 
Similarly, the $\M_n$-moment equations can be written as 
\begin{align}\label{Mnp_mom}
\pder{\M_n'}{\tau} = -\frac{1}{\tau} \big[& u_n\, \L_0 + v_n\, \L_1' + a_n' \M_n' + b_n' \M_{n-1}' 
\nn
&+ c_n' \M_{n+1}' \big] - \frac{\M_n'}{\tR} , \quad \forall n \geq 0,
\end{align}
where $\M_n'\equiv \M_n-\M_n^{\rm eq}$, and  
\begin{align}
u_n &= a_n' \frac{\M_n^{\rm eq}}{\L_0} + b_n' \frac{\M_{n-1}^{\rm eq}}{\L_0} + c_n' \frac{\M_{n+1}^{\rm eq}}{\L_0} -\bar{a}_0 \pder{\M_n^{\rm eq}}{\varepsilon} ,
\nn
v_n &= -c_0  \pder{\M_n^{\rm eq}}{\varepsilon}.
\end{align}

We assume that $\L_n'$ and $\M_n'$ admit a gradient expansion, i.e., an expansion in $\tau_R/\tau=1/w$, starting at order $1/w$. For the $\L_n'$ moments, we write
\begin{equation}\label{Ln_GE}
\frac{\L_n'}{\L_n^{\rm eq}} = \sum_{k=1}^{\infty} \frac{\alpha_{n}^{(k)}}{w^{k}} .
\end{equation}
We note that $\alpha_{n}^{(k)}$ is a function of $m/T$. Using this expansion in Eq.~\eqref{Lnp_mom} and collecting terms of $\mathcal{O}(1/w)$ and $\mathcal{O}(1/w^2)$, we get
\begin{align}\label{Ln_GE_coeff}
\alpha_{1}^{(1)} &= -s_1\frac{\varepsilon}{\L_1^{\rm eq}}\, ,
\nn
\alpha_{1}^{(2)} &= -\bigg[ (\Delta+\bar{a}_0-a_1)s_1 -c_1 s_2 -t_1 s_1 
\nn
&\qquad\ \ + \bar{a}_0 \left( \frac{\del s_1/\del z}{\del \ln \varepsilon/\del z} \right) \bigg] \frac{\varepsilon}{\L_1^{\rm eq}} \,,
\nn
\alpha_{2}^{(1)} &=  -s_2\frac{\varepsilon}{\L_2^{\rm eq}} \,,
\end{align}
where $\Delta = 1-\frac{d\ln \tR}{d\ln \tau}$%
    \footnote{The analysis of this section holds even for a time-dependent relaxation time.}.
To derive these coefficients, we considered the evolution equations for $\L_0, \L_1'$, and $\L_2'$. Thus to order $1/w^2$, we have
\begin{align}\label{L1p_GE}
\L_1' = - \frac{1}{w} s_1 \varepsilon 
- \frac{1}{w^2} \bigg[&(\Delta+\bar{a}_0-a_1)s_1 -c_1 s_2 -t_1 s_1 
\nn
&+ \bar{a}_0 \left( \frac{\del s_1/\del z}{\del \ln \varepsilon/\del z} \right) \bigg] \varepsilon + \mathcal{O}(1/w^3) .
\end{align}
For the massless case, all $s_n, t_n, u_n, v_n$ vanish except $s_1\mapsto b_1$. Further, $\bar{a}_0 \to a_0$ and $\L_1' \to \L_1$. For constant $\tR$ ($\Delta\to 1$) the gradient expansion reduces then to 
\begin{equation}
\L_1' = - \frac{1}{w} b_1 \varepsilon - \frac{1}{w^2} (1+a_0-a_1)b_1 \varepsilon,
\end{equation}
which agrees with \cite{Blaizot:2019scw}.

Similarly, the gradient expansion of $\M_n'$ can be written as 
\begin{equation}
\frac{\M_n'}{\M_n^{\rm eq}} = \sum_{k=1}^{\infty} \frac{\gamma_{n}^{(k)}}{w^{k}} .
\end{equation}
The coefficients are obtained to be 
\begin{align}
\gamma_{0}^{(1)} &= -u_0\frac{\varepsilon}{\M_0^{\rm eq}}\, ,
\nn
\gamma_{0}^{(2)} &= -\bigg[ (\Delta+\bar{a}_0-a_0')u_0 -c_0' u_1 -s_1 v_0 
\nn
&\qquad\ \ + \bar{a}_0 \left( \frac{\del u_0/\del z}{\del \ln \varepsilon/\del z} \right) \bigg] \frac{\varepsilon}{\M_0^{\rm eq}} \,,
\nn
\gamma_{1}^{(1)} &=  -u_1\frac{\varepsilon}{\M_1^{\rm eq}} \,. 
\end{align}
Therefore, up to second order,
\begin{align}\label{M0p_GE}
\M_0' = - \frac{1}{w} u_0 \varepsilon 
-\frac{1}{w^2} \bigg[& (\Delta+\bar{a}_0-a_0')u_0 -c_0' u_1 -s_1 v_0 
\nn
&+ \bar{a}_0 \left( \frac{\del u_0/\del z}{\del \ln \varepsilon/\del z} \right) \bigg] \varepsilon + \mathcal{O}(1/w^3) .
\end{align}

The gradient series for $\Pi$ and $\phi$ can be obtained from Eqs.~(\ref{L1p_GE}, \ref{M0p_GE}). Since $\Pi = -\frac{1}{3}\M_0'$, we obtain:
\begin{align}\label{Bulk_GE}
\Pi = \frac{1}{w} \left(\frac{u_0}{3}\right) \varepsilon 
+ \frac{1}{w^2} \bigg[& (\Delta+\bar{a}_0-a_0')\frac{u_0}{3} -\frac{c_0' u_1}{3} -\frac{s_1 v_0}{3} 
\nn
&+ \frac{\bar{a}_0}{3} \left( \frac{\del u_0/\del z}{\del \ln \varepsilon/\del z} \right) \bigg] \varepsilon + \mathcal{O}(1/w^3) .
\nn
\end{align}
From the relation $\phi= -\frac{1}{3}\M_0' -c_0 \L_1'$, we obtain
\begin{align}\label{Shear_GE}
\phi &= \frac{1}{w} \left(c_0 s_1 +\frac{u_0}{3} \right) \varepsilon 
\nn
&+ \frac{1}{w^2} \bigg[(\Delta+\bar{a}_0-a_1)\left(c_0 s_1 +\frac{u_0}{3} \right) 
-c_0 c_1 s_2 -c_0 t_1 s_1 
\nn
&-\frac{c_0' u_1}{3} -\frac{s_1 v_0}{3} 
+ \bar{a}_0 \left( \frac{c_0 (\del s_1/\del z) + \frac{1}{3} (\del u_0/\del z)}{\del \ln \varepsilon/\del z} \right) \bigg] \varepsilon 
\nn
&+ \mathcal{O}(1/w^3) .
\end{align}

\vspace*{-.2cm}
\subsection{Second-order hydrodynamics} \label{app:gr2}
\vspace*{-.2cm}

We now turn to the hydrodynamic equations and  assume that $\phi$ and $\Pi$ admit the gradient series
\begin{equation}\label{gradhydro}
\frac{\phi}{\varepsilon} =  \sum_{k=1}^{\infty} \frac{\mathcal{S}_k}{w^{k}}, \qquad
\frac{\Pi}{\varepsilon} = \sum_{k=1}^{\infty} \frac{\mathcal{B}_k}{w^{k}} .
\end{equation}
The coefficients $\mathcal{S}_k, \mathcal{B}_k$ are dimensionless functions of $m/T$. These can be determined by plugging the expansions (\ref{gradhydro}) into  the hydrodynamic equations and identifying the various orders in $1/w$. One gets
\begin{align}\label{hydro_GE_coeff}
\mathcal{S}_1 =& \frac{4}{3}\frac{\beta_\phi}{\varepsilon},
\nn
\mathcal{S}_2 =& \bigg[\frac{4}{3} \left(\Delta-\delta_{\phi\phi}-\frac{1}{3}\tau_{\phi\phi} \right) \frac{\beta_\phi}{\varepsilon} -\frac{2}{3} \lambda_{\phi\Pi} \frac{\beta_\Pi}{\varepsilon} 
\nn
&\ + \frac{4}{3} \left( \frac{\bar{a}_0}{\del\ln \varepsilon/\del z} \right) \frac{1}{\varepsilon} \frac{\del \beta_\phi}{\del z} \bigg]  ,
\nn
\mathcal{B}_1 =& -\frac{\beta_\Pi}{\varepsilon},
\nn
\mathcal{B}_2 =& \left[-\left(\Delta -\delta_{\Pi\Pi} \right) \frac{\beta_\Pi}{\varepsilon} +\frac{4}{3} \lambda_{\Pi\phi} \frac{\beta_\phi}{\varepsilon} - \left( \frac{\bar{a}_0}{\del\ln \varepsilon/\del z} \right)  \frac{1}{\varepsilon} \frac{\del \beta_\Pi}{\del z} \right] .
\end{align}
With these coefficients, one can then calculate the gradient expansions of $\phi$ and $\Pi$. One gets
\begin{align}\label{hydro_GE}
\phi =&  \frac{1}{w} \frac{4\beta_\phi}{3} 
+ \frac{1}{w^2} \bigg[\frac{4}{3} \left(\Delta-\delta_{\phi\phi}-\frac{1}{3}\tau_{\phi\phi} \right) \beta_\phi -\frac{2}{3} \lambda_{\phi\Pi} \beta_\Pi 
\nn
&\qquad\qquad\qquad+ \frac{4}{3} \left( \frac{\bar{a}_0}{\del\ln \varepsilon/\del z} \right) \frac{\del \beta_\phi}{\del z} \bigg] + \mathcal{O}(1/w^3) ,
\nn
\Pi =& - \frac{1}{w} \beta_\Pi + \frac{1}{w^2} \bigg[-\left(\Delta -\delta_{\Pi\Pi} \right) \beta_\Pi +\frac{4}{3} \lambda_{\Pi\phi} \beta_\phi 
\nn
&\qquad\qquad\qquad\ \ - \left( \frac{\bar{a}_0}{\del\ln \varepsilon/\del z} \right) \frac{\del \beta_\Pi}{\del z} \bigg] + \mathcal{O}(1/w^3) .
\end{align}

We have checked (numerically for a large range in $m/T$) that the expansions of $\phi$ and $\Pi$ given in  Eqs.~(\ref{Bulk_GE}, \ref{Shear_GE}) and  Eq.~\eqref{hydro_GE} coincide, whatever choice one makes for the transport coefficients  $\delta_{\phi\phi}+\tau_{\phi\phi}/3$ and $\lambda_{\phi\Pi}$  (the coefficients $\lambda_{\Pi\phi}$ and $\delta_{\Pi\Pi}$ being the same in the two cases considered in Sect.~\ref{sec:newcoeff}). We have indeed
\begin{equation}
\frac{4}{3}(\overline{\delta_{\phi\phi}+\tau_{\phi\phi}/3})  \beta_\phi + \frac{2}{3} \overline{\lambda_{\phi\Pi}} \beta_\Pi = \frac{4}{3}(\delta_{\phi\phi}+\tau_{\phi\phi}/3)  \beta_\phi + \frac{2}{3} \lambda_{\phi\Pi} \beta_\Pi \,.
\end{equation}


\bibliography{reference.bib}
\end{document}